\documentclass[prb,reprint,a4paper,superscriptaddress,showpacs,citeautoscript,floatfix]{revtex4-1}

\pdfoutput=1						
\usepackage[charter]{mathdesign}	
\usepackage{amsmath}				

\usepackage[pdftex]{graphicx}\graphicspath{{figs/}}
\usepackage{microtype}
\usepackage{bm}\let\vec\bm

\usepackage[svgnames]{xcolor}
\usepackage[
	colorlinks=True,linkcolor=DarkRed,citecolor=ForestGreen,urlcolor=MediumBlue,
	pdfstartview=FitH,bookmarks=False,pdfpagemode=UseNone
]{hyperref}

\def\Ek{E_{\vec{k}}}
\def\kT{k_{\text{B}}T}
\def\T0{T_0}
\def\kTo{k_{\text{B}}\T0}
\def\tauqp{\tau_{\text{qp}}}
\def\tauopt{\tau_{\text{opt}}}
\def\tauD{\tau_{\text{D}}}
\def\omegaD{\omega_{\text{D}}}
\def\sdc{\sigma_{\text{dc}}}
\def\Zopt{\mathcal{D}}
\def\S{\mathscr{S}}
\def\TFL{T_{\text{FL}}}
\def\omegabar{\bar{\omega}}
\def\Tbar{\bar{T}}
\def\omegae{\omega_{+}}
\def\omegah{\omega_{-}}
\def\omegaL{\omega_{\text{L}}}
\def\omegaH{\omega_{\text{H}}}
\def\PVint{\mathscr{P}\hspace{-1.1em}\int}

\begin{document}

\title{Non-Drude universal scaling laws for the optical response of local Fermi
liquids}

\author{Christophe Berthod}
\affiliation{D{\'e}partement de Physique de la Mati{\`e}re Condens{\'e}e,
Universit{\'e} de Gen{\`e}ve, 24 quai Ernest-Ansermet, 1211 Gen{\`e}ve 4,
Switzerland}

\author{Jernej Mravlje}
\affiliation{Coll{\`e}ge de France, 11 place Marcelin Berthelot, 75005 Paris,
France}
\affiliation{Centre de Physique Th{\'e}orique, {\'E}cole Polytechnique, CNRS,
91128 Palaiseau, France}
\affiliation{Jo\v{z}ef Stefan Institute, Jamova~39, Ljubljana, Slovenia}

\author{Xiaoyu Deng}
\affiliation{Department of Physics, Rutgers University, Piscataway, NJ 08854,
USA}

\author{Rok \v{Z}itko}
\affiliation{Jo\v{z}ef Stefan Institute, Jamova~39, Ljubljana, Slovenia}

\author{Dirk van der Marel}
\affiliation{D{\'e}partement de Physique de la Mati{\`e}re Condens{\'e}e,
Universit{\'e} de Gen{\`e}ve, 24 quai Ernest-Ansermet, 1211 Gen{\`e}ve 4,
Switzerland}

\author{Antoine Georges}
\affiliation{Centre de Physique Th{\'e}orique, {\'E}cole Polytechnique, CNRS,
91128 Palaiseau, France}
\affiliation{Coll{\`e}ge de France, 11 place Marcelin Berthelot, 75005 Paris,
France}
\affiliation{D{\'e}partement de Physique de la Mati{\`e}re Condens{\'e}e,
Universit{\'e} de Gen{\`e}ve, 24 quai Ernest-Ansermet, 1211 Gen{\`e}ve 4,
Switzerland}

\date{26 December 2012}

\begin{abstract}

We investigate the frequency and temperature dependence of the low-energy
electron dynamics in a Landau Fermi liquid with a local self-energy. We show
that the frequency and temperature dependencies of the optical conductivity obey
universal scaling forms, for which explicit analytical expressions are obtained.
For the optical conductivity and the associated memory function, we obtain a
number of surprising features that differ qualitatively from the Drude model and
are universal characteristics of a Fermi liquid. Different physical regimes of
scaling are identified, with marked non-Drude features in the regime where
$\hbar\omega\sim\kT$. These analytical results for the optical conductivity are
compared to numerical calculations for the doped Hubbard model within dynamical
mean-field theory. For the ``universal'' low-energy electrodynamics, we obtain
perfect agreement between numerical calculations and analytical scaling laws.
Both results show that the optical conductivity displays a non-Drude ``foot'',
which could be easily mistaken as a signature of breakdown of the Fermi liquid,
while it actually is a striking signature of its applicability. The
aforementioned scaling laws provide a quantitative tool for the experimental
identification and analysis of the Fermi-liquid state using optical
spectroscopy, and a powerful method for the identification of alternative states
of matter, when applicable.

\end{abstract}

\maketitle

\section{Introduction}
\label{sec:Introduction}

A century after its elaboration, the Drude formula is still commonly used to
analyze the electrodynamic response of metals \cite{Drude-1900, Dressel-2006}.
However, the conceptual basis underlying Drude's phenomenological description
has been entirely changed by the modern quantum theory of the solid state. The
key point is the emergence of long-lived quasiparticle excitations at low energy
and low temperature, which are the actual charge carriers in good metals.
Understanding the emergence of quasiparticles from a correlated liquid of
interacting electrons is the great achievement of Landau's Fermi liquid (FL)
theory\cite{*[] [{ [Sov. Phys. JETP \textbf{3}, 920 (1957)].}] Landau-1956},
which is precisely half as old as the Drude theory. In FL theory, the existence
of long-lived quasiparticles is due to the vanishing of their scattering rate as
the Fermi surface is approached, because of phase-space constraints and the
Pauli principle. For quasiparticles to be well defined, however, they must have
a relaxation rate smaller than their energy (and hence than the available
thermal excitation energy $\sim\kT$), a condition which is met at low
temperatures close to the Fermi surface of simple metals.

When exciting carriers at a low frequency $\omega$, one induces intra-band
transitions of energy $\hbar\omega$ between states within windows of order
$\kT$, thus probing the relaxation rate up to energies, typically, of
$\hbar\omega+\kT$. A Drude-like response is expected if the relaxation rate does
not vary appreciably over this energy range. For Landau quasiparticles in an FL,
the relaxation rate increases quadratically with increasing energy and
temperature. The conditions for a Drude response are therefore met if
$\hbar\omega\ll\kT$, but deviations are expected if $\hbar\omega\gtrsim\kT$.
Deviations from a pure Drude behavior are actually common in metals, and this is
often somewhat loosely interpreted as a violation of Fermi-liquid behavior.
However, in order to distinguish non-Drude features that are consistent
with---or even signatures of---FL behavior from those that indicate a genuine
breakdown of FL theory, one needs to understand the response of Landau
quasiparticles to electromagnetic waves. This is a difficult task in general,
which has regained interest recently \cite{Chubukov-2012, Maslov-2012}, along
with the development of low-frequency spectrometers \cite{Nuss-1998,
Dexheimer-2007}.

In this paper, we address this problem in the context of local Fermi liquids. By
``local'', we mean that the scattering rate, and more generally the
single-particle self-energy, is independent of momentum. From a theory
viewpoint, local FLs are realized, e.g., in the dynamical mean field theory
(DMFT) of strongly correlated electron models. The assumption of a local
self-energy becomes increasingly accurate with increasing coordination number.
It is exact in the limit of infinite dimensionality \cite{Metzner-1989,
Muller-Hartmann-1989}. The success of this approach in practice also
demonstrates that a weak momentum dependence is a reasonable approximation for a
wide class of correlated electron materials, at least in some range of
composition, temperature, etc..., when spatial correlations are short ranged.

The considerable simplification resulting from locality allows us to derive here
analytically a universal scaling form of the optical conductivity in the FL
regime. We show that a Drude-like behavior is indeed recovered in the regime
$\hbar\omega\ll\kT$. In contrast, clear departure from Drude behavior is found
at higher frequency $\hbar\omega\sim\kT$ when the frequency dependence of the
scattering rate becomes important. Characteristic non-Drude signatures of FL
behavior in the optical conductivity are identified in this regime.

We also perform DMFT calculations of the optical conductivity for a microscopic
model of a hole-doped Mott insulator. We show that the results accurately obey
the FL scaling expressions, and that characteristic non-Drude features of the
DMFT optical conductivity in the thermal regime $\hbar\omega\sim\kT$ are
explained by the FL scaling analysis. The model calculation also allows us to
clearly identify the limitations and range of applicability of universal FL
behavior.

Finally, we discuss the conditions for a possible experimental observation of
the FL universal scaling laws and FL signatures in optical measurements.

The paper is organized as follows. Section~\ref{sec:Preliminaries} reviews the
Drude theory and its generalizations. In Sec.~\ref{sec:OpticsFL}, we give the
general formula for the optical conductivity of a system with a local
self-energy. We then introduce the low-energy expression of the self-energy in a
local FL, derive analytically a universal scaling expression for the optical
conductivity, and discuss the different regimes of physical relevance.
Section~\ref{sec:OpticsDMFT} presents a comparison of the FL scaling laws with
DMFT calculations. In Sec.~\ref{sec:Experiment}, we address issues related to
the experimental observation of FL scaling laws in the optical conductivity, and
discuss in more detail the case of UPd$_2$Al$_3$. Our conclusions are given in
Sec.~\ref{sec:Conclusion}, and a series of appendices collect additional
material.

\section{Preliminaries -- Drude theory and beyond}
\label{sec:Preliminaries}

In the Drude theory of conduction in metals, a single frequency-independent time
$\tauD$ governs the relaxation of the current. The assumption is that the
current decays exponentially once the electric field has been turned off. The
classical equation of motion then leads to the dc conductivity
$\sdc=ne^2\tauD/m$, with $n$ the carrier density and $m$ the carrier mass. In
the presence of an oscillating electric field, the complex frequency-dependent
conductivity $\sigma(\omega)=\sigma_1(\omega)+i\sigma_2(\omega)$ reads:
	\begin{equation}\label{eq:Drude_conductivity}
		\sigma(\omega)=\frac{ne^2}{m}\frac{1}{-i\omega+1/\tauD}
		=\frac{\sdc}{1-i\omega\tauD}.
	\end{equation}
In many materials, however, especially those with strong electron correlations,
a single frequency-independent relaxation time does not provide an accurate
description of the experimental data \cite{Basov-2011}. In order to describe the
full frequency dependence of the conductivity, it is convenient to introduce
\cite{Gotze-1972} a memory function $M(\omega)$ such that:
	\begin{equation}\label{eq:conductivity_memory}
		\sigma(\omega)=\frac{i\epsilon_0\omega_p^2}{\omega+M(\omega)}.
	\end{equation}
In this expression, we define the plasma frequency from the sum rule over the
whole frequency range:
	\begin{equation}\label{eq:def_omegap}
		\epsilon_0\omega_p^2\equiv
		\frac{2}{\pi}\int_0^{\infty}d\omega\,\sigma_1(\omega).
	\end{equation}
The complex function $M(\omega)=M_1(\omega)+iM_2(\omega)$ plays the role of a
self-energy for the optical response. The definition (\ref{eq:def_omegap}) of
$\omega_p^2$ ensures that $\omega$ dominates over $M(\omega)$ in the expression
of $\sigma(\omega)$ at large frequencies. Indeed, the Kramers-Kronig relations
	\begin{equation}\label{eq:KK}
		\{\sigma_1(\omega),\,\sigma_2(\omega)\} = 
		\frac{1}{\pi}\PVint_{-\infty}^{\infty}d\Omega\,
		\frac{\{-\sigma_2(\Omega),\,\sigma_1(\Omega)\}}{\omega-\Omega}
	\end{equation}
imply that $\sigma_2(\omega\sim\infty)\sim 2\int_0^\infty
d\Omega\,\sigma_1(\Omega)/(\pi\omega)$.

Expression (\ref{eq:conductivity_memory}) can be cast in a form that is
formally analogous to the Drude expression (``generalized Drude model''):
	\begin{equation}\label{eq:gen_Drude}
		\sigma(\omega)=\epsilon_0\omega_p^2
		\frac{\Zopt(\omega)}{-i\omega+1/\tauopt(\omega)}
	\end{equation}
with:
	\begin{equation}
		\Zopt(\omega)=\left[1+\frac{M_1(\omega)}{\omega}\right]^{-1},\quad
		\frac{1}{\tauopt(\omega)}=\Zopt(\omega) M_2(\omega).
	\end{equation}
Note that the complex conductivity obeys $\sigma^*(\omega)=\sigma(-\omega)$
under complex conjugation, so that $\sigma_1$, $M_2$, $\Zopt$ (respectively,
$\sigma_2$, $M_1$) are even (respectively, odd) functions of frequency.
$\Zopt(\omega)$ is often denoted $m/m^*(\omega)$, hence defining an optical
effective mass, and $\tauopt(\omega)$ is often written as $\tau^*(\omega)$. At
low frequency, $\Zopt(\omega\rightarrow 0)=\left(1+\partial_{\omega}
M_1|_{\omega=0}\right)^{-1}$ renormalizes the bare plasma frequency and the
spectral weight of the Drude peak. In many cases, the frequency dependence of
the optical scattering rate $1/\tauopt(\omega)$ at low frequency is mainly
determined by $M_2(\omega)$, with $\Zopt(\omega)$ having a milder frequency
dependence (see below).

When analyzing experimental data, the imaginary part of the memory function,
which controls the optical scattering rate, can, for example, be determined
through:
	\begin{equation}\label{eq:M2}
		M_2(\omega)=\text{Re}\,\frac{\epsilon_0\omega_p^2}{\sigma(\omega)},
	\end{equation}
with the plasma frequency determined by the sum rule (\ref{eq:def_omegap}).

\section{Optical conductivity of local Fermi liquids}
\label{sec:OpticsFL}

We now specialize the discussion to the optical conductivity of \emph{local}
Fermi liquids, i.e., systems in which the single-particle self-energy $\Sigma$
obeys the low-frequency, low-temperature behavior of Landau Fermi-liquid theory
and, additionally, does not depend on momentum. This is the case, in particular,
of strongly correlated electron models and materials treated in the framework of
dynamical mean-field theory \cite{Georges-1992}.

We first recall the simplifications encountered in the Kubo formalism in this
context, and the resulting expression of the optical conductivity. We then show
that universal scaling laws emerge in the Fermi liquid regime and discuss these
laws in the different physical regimes.

\subsection{General expression of the optical conductivity} 

When the single-particle self-energy $\Sigma(\varepsilon)$ has no momentum
dependence, the Kubo formalism leads to the following general expression of the
optical conductivity:
	\begin{subequations}\label{eq:sigma_Phi}
	\begin{multline}\label{eq:sigma_1}
		\sigma_1(\omega)=\frac{1}{\omega}\int_{-\infty}^{\infty}
		d\varepsilon\,[f(\varepsilon)-f(\varepsilon+\hbar\omega)] \\
		\times \pi\int_{-\infty}^{\infty} d\xi\,\Phi(\xi)\,A(\xi,\varepsilon)
		A(\xi,\varepsilon+\hbar\omega),
	\end{multline}
where $f(\varepsilon)$ is the Fermi function, while $\Phi(\xi)$ and
$A(\xi,\varepsilon)$ are, respectively, the transport and the
one-particle spectral functions, defined in detail below. The derivation of this
formula is outlined in Appendix~\ref{app:Kubo}. The key point is that vertex
corrections associated with the current-current correlation function exactly
vanish in the case of a local (momentum independent) theory \cite{Khurana-1990},
so that the conductivity can be entirely expressed in terms of the one-particle
self-energy.

Expression (\ref{eq:sigma_1}) of the optical conductivity applies to a
single-band system, to which our discussion is limited in this paper for
simplicity. The entire information about the band dispersion is encoded in the
transport function $\Phi$, defined by:
	\begin{equation}\label{eq:Phi_xi}
		\Phi(\xi)=\frac{2e^2}{dL^d}\sum_{\vec{k}}v^2_{\vec{k}}
		\delta(\xi-\xi_{\vec{k}}).
	\end{equation}
In this expression, $L$ is the system size, $d$ is the dimensionality,
$\xi_{\vec{k}}=\Ek-\mu$ is the dispersion of the non-interacting Bloch band
measured from the chemical potential, and
$\vec{v}_{\vec{k}}=(1/\hbar)\vec{\nabla}_{\vec{k}}\Ek$ is the corresponding
group velocity. Many-body effects enter through the single-electron spectral
function, $A(\xi,\varepsilon)$, which is related to the self-energy
$\Sigma(\varepsilon)=\Sigma_1(\varepsilon)+i\Sigma_2(\varepsilon)$ by
	\begin{equation}
		A(\xi,\varepsilon)=\frac{-\Sigma_2(\varepsilon)/\pi}
		{[\varepsilon-\xi-\Sigma_1(\varepsilon)]^2+[\Sigma_2(\varepsilon)]^2}.
	\end{equation}
	\end{subequations}
In all these expressions, $\varepsilon/\hbar$ designates a frequency, while the
momentum dependence of the spectral function stems from $\xi_{\vec{k}}$.

The transport function $\Phi(\xi)$ is usually a slow function of its argument,
in contrast to $A(\xi,\varepsilon)$ which in a Fermi liquid peaks at
$\varepsilon\approx\xi$. When the energy dependence of $\Phi(\xi)$ is negligible
over the energy range where the spectral functions are appreciable, the second
integral in Eq.~(\ref{eq:sigma_1}) reduces to the convolution of two Lorentzian
functions, and the real and imaginary parts of the complex conductivity can be
recast into the very convenient form \cite{Lee-1989, Shulga-1991, Allen-2004}
(see Appendix~\ref{app:Kubo})
	\begin{equation}\label{eq:sigma_Sigma}
		\sigma(\omega)=\frac{i\Phi(0)}{\omega}\int_{-\infty}^{\infty}
		d\varepsilon\,\frac{f(\varepsilon)-f(\varepsilon+\hbar\omega)}
		{\hbar\omega+\Sigma^*(\varepsilon)-\Sigma(\varepsilon+\hbar\omega)}.
	\end{equation}
Equation (\ref{eq:sigma_Sigma}) reduces to the Drude formula,
Eq.~(\ref{eq:Drude_conductivity}), if we put
$\Sigma(\varepsilon)\equiv-i\hbar/(2\tauD)$, and if $\Phi(0)$ is evaluated using
the three-dimensional isotropic electron gas formula.

In the context of a single-band system, the total sum rule as defined from
Eq.~(\ref{eq:def_omegap}) reads (see Appendix~\ref{app:sum-rule}):
	\begin{equation}\label{eq:sumrule_1band}
		\epsilon_0\omega_p^2=\int_{-\infty}^{\infty} d\xi\,\Phi(\xi)
		\left(-\frac{dn}{d\xi}\right)
		=\frac{2e^2}{d\hbar^2}\langle n(\xi_{\vec{k}})\nabla^2E_{\vec{k}}
		\rangle_{\text{BZ}}
	\end{equation}
where $\langle\cdots\rangle_{\text{BZ}}$ designates an average over the
Brillouin zone, and $n(\xi_{\vec{k}})=\langle
c^\dagger_{\vec{k}}c^{\phantom{\dagger}}_{\vec{k}}\rangle=\int d\varepsilon\,
f(\varepsilon)\,A(\vec{k},\varepsilon)$ is the momentum distribution of the
electrons. In an interacting Fermi liquid, $-dn/d\xi$ is different from a
$\delta$ function even at zero temperature, hence $\epsilon_0\omega_p^2$ is, in
general, different from $\Phi(0)$. For a simple tight-binding band
$\Ek\propto\sum_\alpha\cos k_\alpha$, the right-hand side becomes proportional
to the kinetic energy. Note that the value of the sum rule thus depends on
temperature and interaction strength. This is in contrast to the case where the
whole solid (with all bands) is considered, in which case the plasma frequency
$\omega_p^2$ is set by the bare electron mass and total number of electrons
($f$-sum rule). Note also that the approximate formula (\ref{eq:sigma_Sigma}),
in which the transport function was replaced by a constant, does not reproduce
correctly the total sum rule, which is expected since it is only valid at low
energy. We shall see, however, that it can, in general, be used to reliably
estimate the spectral weight of the Drude peak.

\subsection{Scattering in a local Fermi liquid}

At low frequency and temperature, the self-energy of a local Fermi liquid can be
written in the form:
	\begin{equation}\label{eq:self-energy}
		\Sigma(\varepsilon,T)=\left(1-\frac{1}{Z}\right)\varepsilon
		-\frac{i}{Z\pi\kTo}\left[\varepsilon^2+(\pi\kT)^2\right].
	\end{equation}
In this expression, $Z$ is the quasiparticle spectral weight. In a local Fermi
liquid, it is also related to the quasiparticle mass renormalization through
$Z=m/m^*$ (i.e., the quasiparticle Fermi velocity is
$v_{\text{F}}^*=Zv_{\text{F}}$). Close to the Fermi surface and for
$\varepsilon\ll\kT$, the spectral function is approximately a Lorentzian peak of
weight $Z$, centered around $\varepsilon=Z\xi_{\vec{k}}$ (the quasiparticle
dispersion). We define the temperature-dependent quasiparticle lifetime from the
zero-frequency value of $\Sigma_2$ as:
	\begin{equation}\label{eq:def_tauqp}
		\frac{\hbar}{\tauqp} \equiv 2Z|\Sigma_2(\varepsilon=0,T)|
		=2\pi\frac{(\kT)^2}{\kTo}.
	\end{equation}
It corresponds to twice the width of the Lorentzian quasiparticle peak in the
spectral function, that is, to the decay rate of the probability (square of the
Green's function).

\begin{figure}[tb]
\includegraphics[width=\columnwidth]{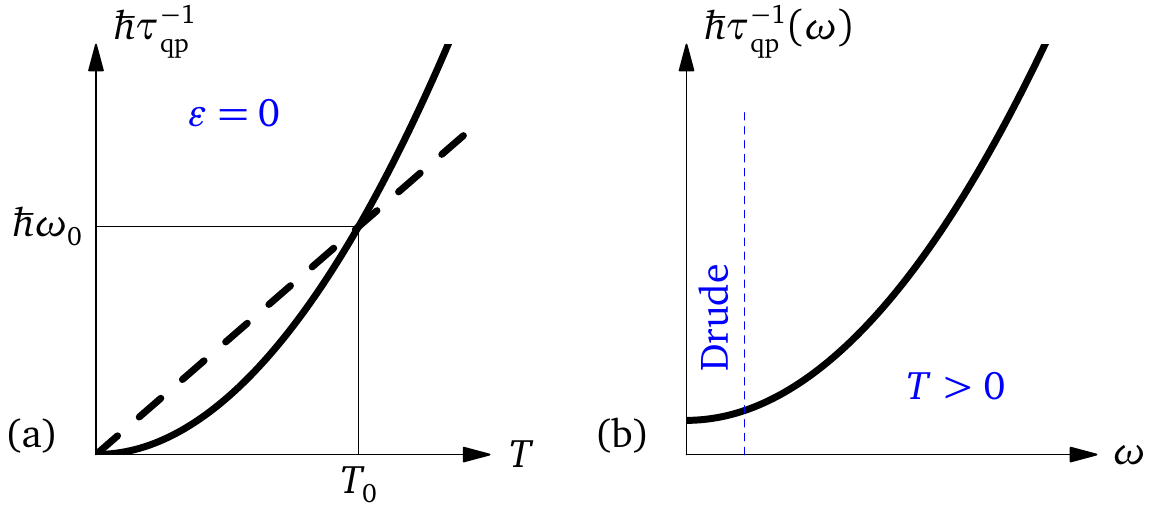}
\caption{\label{fig:scattering}
Fermi liquid scattering rate $\hbar\tauqp^{-1}$. (a) At zero energy, $\hbar
\tauqp^{-1}$ increases quadratically with temperature (solid line). The dashed
line indicates $2\pi\kT$. The intercept defines a temperature scale $\T0$ and a
frequency scale $\hbar\omega_0=2\pi\kTo$. (b) At finite $T$,
$\hbar\tauqp^{-1}(\omega)$ increases quadratically with the frequency $\omega$.
The applicability of the frequency-independent (Drude) approximation is limited
to low frequencies.
}
\end{figure}

Fermi liquid behavior self-consistently relies on the existence of a sharp Fermi
surface, and applies when the scattering rate of the typical excitation is
smaller than its energy. The scattering rate $\hbar/\tauqp$ must be compared to
a typical thermal excitation energy, say $2\pi\kT$. The characteristic Fermi
liquid energy scale $\kTo$ is defined as the temperature where
$\hbar/\tauqp=2\pi\kT$, and will serve as a basic unit below. This is
illustrated in Fig.~\ref{fig:scattering} (left panel), which sketches the
dependence of the quasiparticle lifetime on temperature. The temperature $\T0$
also sets the coupling strength: a small $\T0$ corresponds to strong
electron-electron interactions and a large value of $\Sigma_2$. Hence the
quasiparticle lifetime diminishes rapidly as a function of temperature and
frequency. The precise relation of the scale $\T0$ to the scale $\TFL$, below
which strict Fermi liquid behavior holds, is discussed in
Sec.~\ref{sec:OpticsDMFT}. In strongly correlated (local) Fermi liquids in which
$Z$ is small, e.g., close to a Mott transition (see Sec.~\ref{sec:OpticsDMFT})
or in heavy-fermion materials, all physical quantities scale with a single
energy scale, and $\T0$ is proportional to $\sim ZD$ (with possibly a small
value of the prefactor). Here, $D$ is the half-bandwidth of the bare band. The
factor $1/Z$ which has been pulled out in front of $\Sigma_2$ in order to define
$\T0$ ensures that in such a case $\Sigma_2$ is a scaling function of $T/ZD$ and
$\varepsilon/ZD$.

The right panel of Fig.~\ref{fig:scattering} suggests that the frequency
dependence of the quasiparticle scattering rate is actually important. A
Drude-like optical response with a constant relaxation time $\tauD\sim\tauqp$
can only be expected to provide a reasonable description in the very
low-frequency or relatively high-temperature regime
$\hbar\omega\lesssim2\pi\kT$. When $\hbar\omega\gtrsim2\pi\kT$, the energy
dependence of $\Sigma(\varepsilon)$ cannot be neglected and a non-Drude response
arises, as discussed in the following section.

We finally note that a real, frequency-independent Hartree term $\Sigma_1(0,T)$
should in fact be added to Eq.~(\ref{eq:self-energy}) for completeness. It sets
the location of the Fermi surface from $\Ek=\mu-\Sigma_1(0,T)$ and can be viewed
as a shift of the chemical potential, which will be omitted for simplicity in
all equations. An imaginary frequency-independent part can also be added to
mimic the effects of the impurity scattering. This is considered in
Appendix~\ref{app:impurity_scattering}.

\subsection{Scaling form of the optical conductivity in a local Fermi liquid}

We now show that the optical conductivity obeys a universal scaling form in
terms of the two variables $\omega\tauqp$ and $\hbar\omega/(2\pi\kT)$. Whereas
it reduces essentially to the Drude form in the low-frequency limit, its full
frequency dependence is markedly different.

The universal scaling form is derived by inserting Eq.~(\ref{eq:self-energy})
into Eq.~(\ref{eq:sigma_Sigma}). The calculations can be performed analytically
and yield:
	\begin{subequations}\label{eq:Fermi_liquid}
	\begin{align}
		\label{eq:Fermi_liquid_a}
		\sigma(\omega)&=\sdc\,\S\left(\frac{\hbar\omega}{2\pi\kT},
		\omega\tauqp\right),\\
		\sdc&=\frac{\pi^2}{12}Z\Phi(0)\tauqp,\\
		\nonumber
		\S(x,y)&=\frac{6}{\pi^2x}\int_{-\infty}^{\infty} du\,
		\frac{[e^{\pi(u-x)}+1]^{-1}-[e^{\pi(u+x)}+1]^{-1}}{1+x^2-iy+u^2}.
	\end{align}
The scaling function $\S$ is evaluated and displayed in
Appendix~\ref{app:S}. One obtains:
	\begin{multline}\label{eq:S_function}
		\S(x,y)=\frac{6i}{\pi^2}\frac{1}{x\,r(x,y)}\left\{
		\psi\left(\textstyle\frac{1}{2}\left[1+r(x,y)-ix\right]\right)
		\right.\\\left.
		-\psi\left(\textstyle\frac{1}{2}\left[1+r(x,y)+ix\right]\right)\right\},
	\end{multline}
	\end{subequations}
where $r(x,y)=\sqrt{1+x^2-iy}$ and $\psi$ is the digamma function defined as
$\psi(z)=\lim_{M\to\infty}\big[\ln M-\sum_{n=0}^{M}1/(n+z)\big]$.

Equation (\ref{eq:Fermi_liquid_a}) emphasizes the emergence of two natural
time/frequency scales: the quasiparticle time $\tauqp$, and a ``coherence'' time
$\hbar/(2\pi\kT)$. Alternatively, one can reexpress
	\begin{equation}\label{eq:conductivity_dimensionless}
		\sigma(\omega)=\sdc\,\S
		\left(\frac{\omegabar}{\Tbar},\frac{\omegabar}{\Tbar^2}\right)
	\end{equation}
with $\omegabar\equiv\hbar\omega/(2\pi\kTo)$ and $\Tbar\equiv T/\T0$
dimensionless variables normalized to the basic scale $\T0$. This emphasizes
that the optical conductivity in a Fermi liquid is a scaling function of
$\omega/T$ and $\omega/T^2$. Equations (\ref{eq:Fermi_liquid}) constitute the
key analytical result of this article. They replace the Drude formula by a
universal scaling form of these \emph{two} frequency scales, which is valid for
local Fermi liquids. Let us emphasize that all high-energy scales such as the
bare bandwidth or the plasma frequency $\omega_p$ have disappeared from the
scaling expression (\ref{eq:Fermi_liquid}). Instead, only low-energy scales
appear, such as $\T0$ and $Z\Phi(0)$ (the latter is shown below to be related to
the low-energy Drude weight).

\begin{figure}[tb]
\includegraphics[width=0.85\columnwidth]{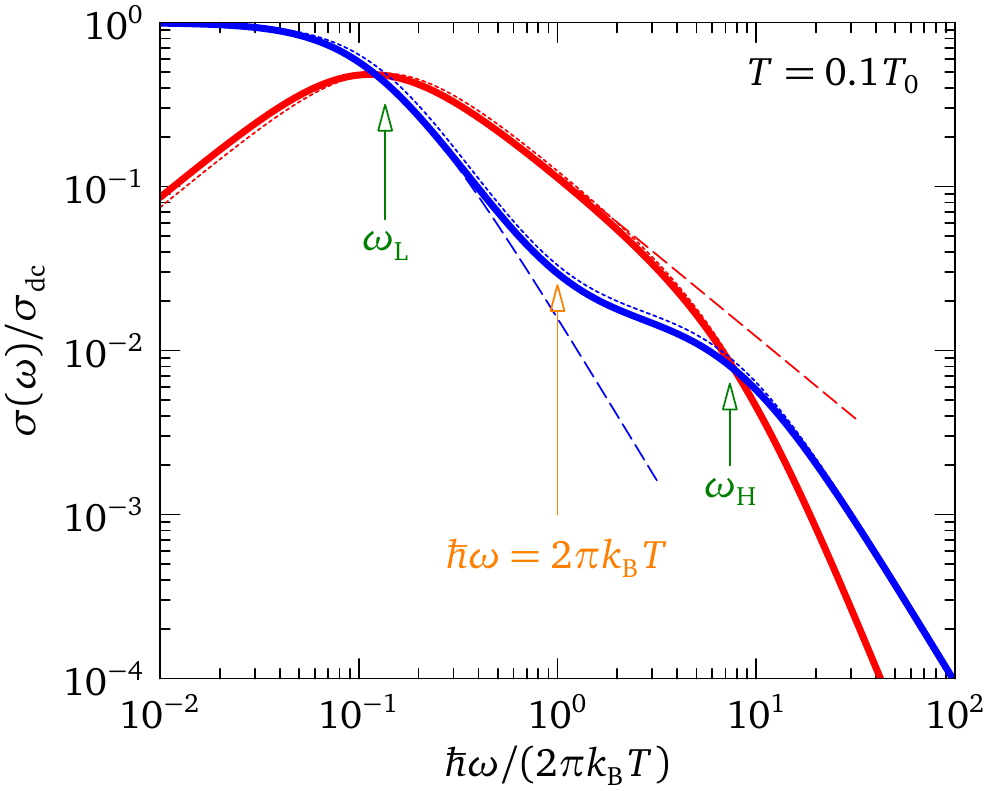}
\caption{\label{fig:log-log-lowT}
Fermi-liquid conductivity Eqs.~(\ref{eq:Fermi_liquid}) at low temperature (solid
lines), below the temperature $T_1$ defined in Fig.~\ref{fig:regimes}. The blue
(red) lines show the real (imaginary) part of the conductivity. The dashed lines
show the low-frequency Drude-like behavior given by Eq.~(\ref{eq:Drude_regime}).
The dotted lines show Eq.~(\ref{eq:sigma-tauopt}). The characteristic frequency
scales $\omegaL$ and $\omegaH$ correspond to those defined in
Fig.~\ref{fig:regimes} and Eq.~(\ref{eq:omegapm}).
}
\end{figure}

In Fig.~\ref{fig:log-log-lowT}, we plot the real and the imaginary parts of
$\sigma(\omega)/\sdc$ as a function of $\hbar\omega/(2\pi\kT)$ on a log-log
scale at a given low temperature $T/\T0=0.1$. (The full frequency and
temperature dependencies are shown on a three-dimensional plot in
Appendix~\ref{app:S}.) Three frequency regimes can be identified from this plot:
\begin{itemize}

\item At low-frequency $\omega\lesssim\omegaL<2\pi\kT/\hbar$ (with $\omegaL$ to
be made precise below, of order $\tauqp^{-1}$ at low temperature), the
conductivity follows closely the Drude model, with a saturation of $\sigma_1$
below the characteristic frequency $\tauqp^{-1}$ and a $1/\omega^2$ decay above
it. $\tauqp^{-1}$ also separates the dissipative regime (with larger $\sigma_1$)
from the inductive regime (with larger $\sigma_2$).

\item When $\omega$ approaches $2\pi\kT/\hbar$, the conductivity deviates from
the Drude behavior: $\sigma_1$ displays a pronounced shoulder with much weaker
frequency dependence---the feature appears as a shoulder in a log-log plot, as a
``foot'' in a lin-lin plot, see below. In this ``thermal'' regime, the
conductivity behaves inductively ($\sigma_2>\sigma_1$) rather than dissipatively.

\item Increasing $\omega$ further, leads to a more rapid decay of $\sigma_2$,
and at $\omegaH$ (to be defined below), the data become dissipative-like again
($\sigma_1/\sigma_2>1$). This is actually a consequence of the assumed unbounded
quadratic increase of the scattering rate with frequency, and might not be
physical in this already high-frequency regime (see Sec.~\ref{sec:OpticsDMFT}).
Above $\omegaH$, $\sigma_1$ recovers a Drude-like $1/\omega^2$ decay, while
$\sigma_2$ turns to $1/\omega^3$. The $1/\omega^3$ behavior is an artifact of
extending the $\omega^2$ in the self-energy (\ref{eq:self-energy}) to high
energies.\footnote{One sees the following from Eq.~(\ref{eq:sigma-tauopt}):
$\sigma_2\sim 1/[\omega+\tauopt^{-2}(\omega)/\omega]$. Hence the $1/\omega^3$
results from the $\omega^2$ of the optical scattering rate. At frequencies well
above $\omega_{\pm}$ (defined in Sec.~\ref{sec:OpticsDMFT}), $\tauopt^{-1}$ is
expected to saturate to a frequency-independent value, so that the correct
asymptotic $1/\omega$ decay of $\sigma_2$ is recovered. The saturation of
$\tauopt^{-1}$ does not affect the power law of $\sigma_1$, since it can be seen
to go like $1/[\omega^2\tauopt(\omega)+\tauopt^{-1}(\omega)]$.}

\end{itemize}

\begin{figure}[tb]
\includegraphics[width=0.85\columnwidth]{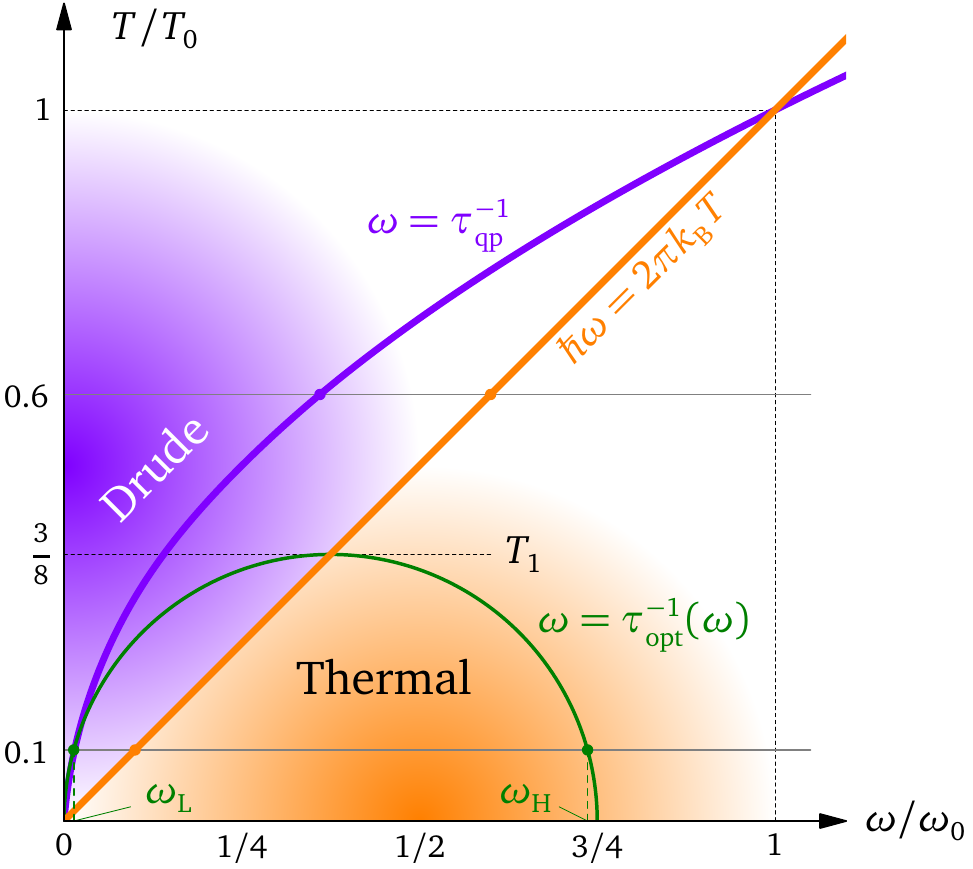}
\caption{\label{fig:regimes}
Regimes of conduction. The Drude regime applies if $\omega<2\pi\kT/\hbar$, and
the thermal regime if $\omega\gtrsim 2\pi\kT/\hbar$. The conductivity along the
two horizontal cuts at fixed temperature is displayed in
Figs.~\ref{fig:log-log-lowT} and \ref{fig:log-log-highT}.
}
\end{figure}
 
These different regimes, as well as the precise conditions determining the
characteristic frequencies $\omega_{\text{L,\,H}}$ are represented on
Fig.~\ref{fig:regimes}. We now discuss these different regimes in more detail.

\subsection{Drude regime} 

The lowest frequency regime can legitimately be called ``Drude regime'', since
there the frequency dependence of the scattering rate can be ignored. In this
regime, the frequency $\hbar\omega$ is small with respect to $2\pi\kT$, the
inverse quasiparticle time $\tauqp^{-1}$ is smaller than, or comparable to the
temperature, but the product $\omega\tauqp\equiv y$ can take arbitrary values
(see Fig.~\ref{fig:regimes}). This regime is thus described by taking the limit
$x\equiv\hbar\omega/(2\pi\kT)\rightarrow 0$ in the scaling form
(\ref{eq:S_function}). This yields:
	\begin{equation}\label{eq:Drude_regime}
		\frac{\sigma(\omega)}{\sdc}\approx\frac{6}{\pi^2}\frac{\psi'
		\left(\textstyle\frac{1}{2}\left[1+\sqrt{1-i\omega\tauqp}\right]\right)}
		{\sqrt{1-i\omega\tauqp}}
	\end{equation}
where $\psi'$ is the derivative of the digamma function. This is shown as the
dashed lines in Fig.~\ref{fig:log-log-lowT}. Interestingly, this expression
differs from the simple Drude form $\sigma(\omega)/\sdc=1/(1-i\omega\tauD)$.
However, as detailed in Appendix~\ref{sec:comparison_Drude}, the frequency
dependence of the exact expression (\ref{eq:Drude_regime}) can be rather
accurately approximated by the simple Drude form, provided $\tauD$ is chosen in
an appropriate manner (which is such that $\tauD$ {\it differs} from $\tauqp$,
however).

The spectral weight in the Drude peak can be estimated as:
	\begin{equation}\label{eq:Drude_weight_main}
		\frac{2}{\pi}\int_0^{\omegaD}d\omega\,\sigma_1(\omega) = Z\Phi(0),
	\end{equation}
in which $\omegaD$ is a cutoff defining the Drude regime. This expression can be
established in two ways. The first is to perform a direct frequency integration
of the scaling expression (\ref{eq:Drude_regime}) over the whole frequency range
(i.e., for the scaling variable $y=\omega\tauqp$ varying from $0$ to $\infty$).
Alternatively, since $\Phi(\xi)$ can be taken as constant in the
low-frequency range, one can use expression (\ref{eq:sigma_Sigma}) and observe
that it yields a high-frequency behavior of $\sigma_2\sim Z\Phi(0)/\omega$. By
Kramers-Kronig, this leads to Eq.~(\ref{eq:Drude_weight_main}).

Two remarks are in order regarding expression (\ref{eq:Drude_weight_main}).
First, the Drude weight scales with the quasiparticle weight $Z$. Close to a
Mott transition, for example, $Z$ is expected to vanish and so does the Drude
weight. Second, it should be noted that $Z$ measures the renormalization of the
Drude weight as compared to the non-interacting (band) value $\Phi(0)$ and
\emph{not} the spectral weight of the Drude peak relative to the total
integrated spectral weight $\epsilon_0\omega_p^2$. The latter relative weight is
given by $Z\Phi(0)/(\epsilon_0\omega_p^2)\approx\Zopt(0)$ with
$\epsilon_0\omega_p^2$ given by Eq.~(\ref{eq:sumrule_1band}).

\subsection{\boldmath Thermal regime and emergence of the $2\pi$ factor in the
optical scattering rate}

In the thermal regime, $\hbar\omega$ is comparable to $2\pi\kT$ and
$\omega\tauqp \gg 1$. This corresponds to the limit of fixed
$x=\hbar\omega/(2\pi\kT)$ and large $y=\omega\tauqp \gg x$. The scaling function
$\S(x,y)$ has the following expansion for $y\to\infty$:
	\begin{equation}\label{eq:Sxyinf}
		\S(x,y\to\infty)=\frac{12i}{\pi^2y}+\frac{16}{\pi^2y^2}(1+x^2).
	\end{equation}
Using this expression and performing a large-$y$ expansion of
$1/\sigma=1/[\sdc\S(x,y)]$, one directly obtains a generalized Drude form
for the optical conductivity:
	\begin{equation}\label{eq:sigma-tauopt}
		\sigma(\omega)\approx\frac{Z\Phi(0)}{-i\omega+1/\tauopt(\omega)},
	\end{equation}
with the optical scattering rate
	\begin{equation}\label{eq:tauopt}
		\frac{\hbar}{\tauopt(\omega)}=\frac{2}{3\pi\kTo}
		\left[(\hbar\omega)^2+(2\pi\kT)^2\right].
	\end{equation}
The same result is found by approximating $\mathscr{D}(\omega)$ in
Eq.~(\ref{eq:gen_Drude}) by $\mathscr{D}(0)\approx
Z\Phi(0)/(\epsilon_0\omega_p^2)$, identifying with Eqs.~(\ref{eq:Fermi_liquid}),
and expanding for large $y$. Equation~(\ref{eq:sigma-tauopt}) is displayed in
Fig.~\ref{fig:log-log-lowT} as the dotted lines. While the Fermi-liquid form of
the quasiparticle lifetime (a one-particle quantity) involves
$(\hbar\omega)^2+(\pi\kT)^2$, it should be emphasized that the optical
scattering rate (a two-particle quantity) involves instead a factor $2\pi$ in
the combination $(\hbar\omega)^2+(2\pi\kT)^2$. This was
emphasized by Gurzhi \cite{*[{See }] [{, and references therein.}] Gurzhi-1959}.
Experimentally, the $2\pi$ factor has not been observed so far in simple metals.
As discussed in detail in Sec.~\ref{sec:Experiment}, in several correlated
metals, a scaling of the optical scattering rate following
$(\hbar\omega)^2+(p\pi\kT)^2$ was reported, with values of $p$ ranging from $1$
to $2.4$ \cite{Sulewski-1988, Katsufuji-1999, Yang-2006, Dressel-2011,
Mirzaei-2012, Nagel-2012}. The departure from the Fermi-liquid value $p=2$ may
be attributed to scattering mechanisms with a different frequency/temperature
dependence \cite{Chubukov-2012, Maslov-2012}. In
Appendix~\ref{app:impurity_scattering}, we show that a frequency-independent
scattering rate does not change the value $p=2$.

\begin{figure}[tb]
\includegraphics[width=0.85\columnwidth]{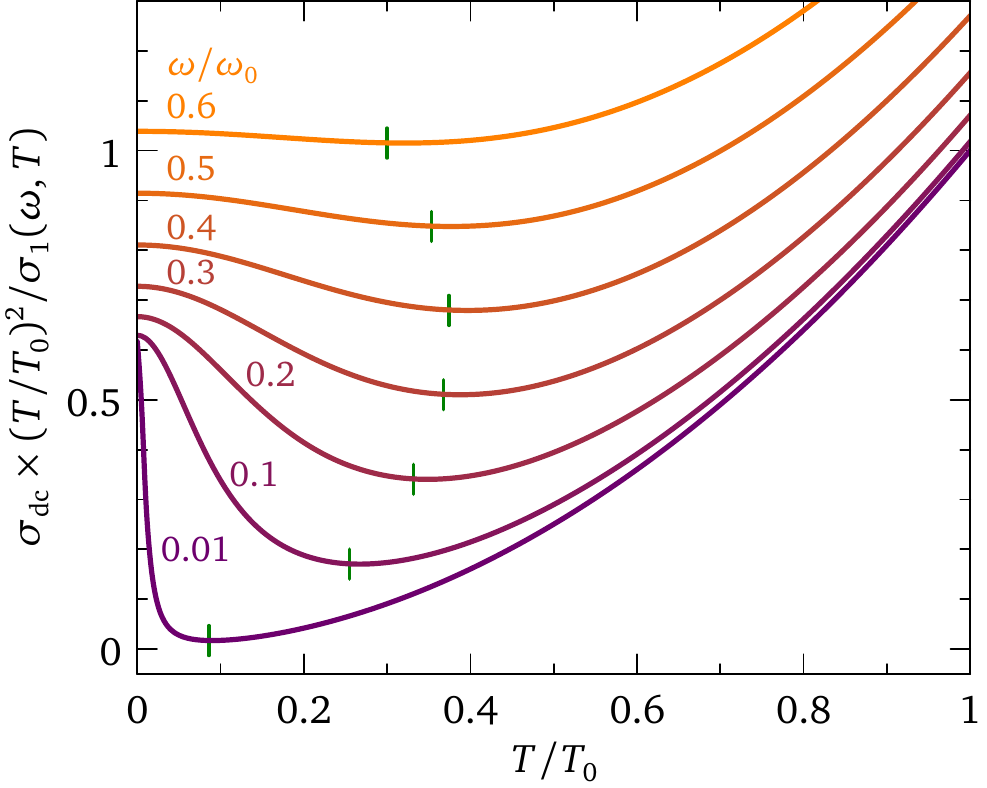}
\caption{\label{fig:minimum}
ac resistivity $1/\sigma_1(\omega,T)$ as a function of $T$ for several
frequencies $\omega$. These curves correspond to vertical slices in
Fig.~\ref{fig:regimes}, and the green bars indicate the temperature at which the
dome is crossed for each frequency. The temperatures are measured in units of
$\T0$, and the frequencies in units of $\omega_0=2\pi\kTo/\hbar$. The
normalization of the resistivity is $\sdc(T/\T0)^2=\pi\hbar Z\Phi(0)/(24\kTo)$.
}
\end{figure}

At low temperature, Eqs.~(\ref{eq:sigma-tauopt}) and (\ref{eq:tauopt}) allow to
define two characteristic frequencies $\omega_{\text{L,\,H}}$ from the condition
$\omega=\tauopt^{-1}(\omega)$, or equivalently
$\sigma_1(\omega)=\sigma_2(\omega)$:
	\begin{equation}\label{eq:omegapm}
		\hbar\omega_{\text{L,\,H}}=\frac{3}{4}\pi\kTo
		\left[1\pm\sqrt{1-\left(\frac{8T}{3\T0}\right)^2}\right].
	\end{equation}
Plotted as a function of $T$, these two frequencies form a dome in the
$(\omega,T)$ plane, defining the thermal regime (green line in
Fig.~\ref{fig:regimes}). Below this dome, the conductivity behaves inductively.
While the scales $\omega_{\text{L,\,H}}$ can be determined from the crossings of
$\sigma_1$ and $\sigma_2$ plotted as a function of frequency
(see Fig.~\ref{fig:log-log-lowT}), the crossing of the dome is most easily
identified by a minimum in a plot of $1/\sigma_1(\omega,T)$ as a function of
$T$, at a fixed finite frequency. This is illustrated in Fig.~\ref{fig:minimum}.
As this determination of $\omega_{\text{L,\,H}}$ only relies on $\sigma_1$, it
might be the most direct way of checking FL behavior in experimental datasets.

\begin{figure}[b]
\includegraphics[width=\columnwidth]{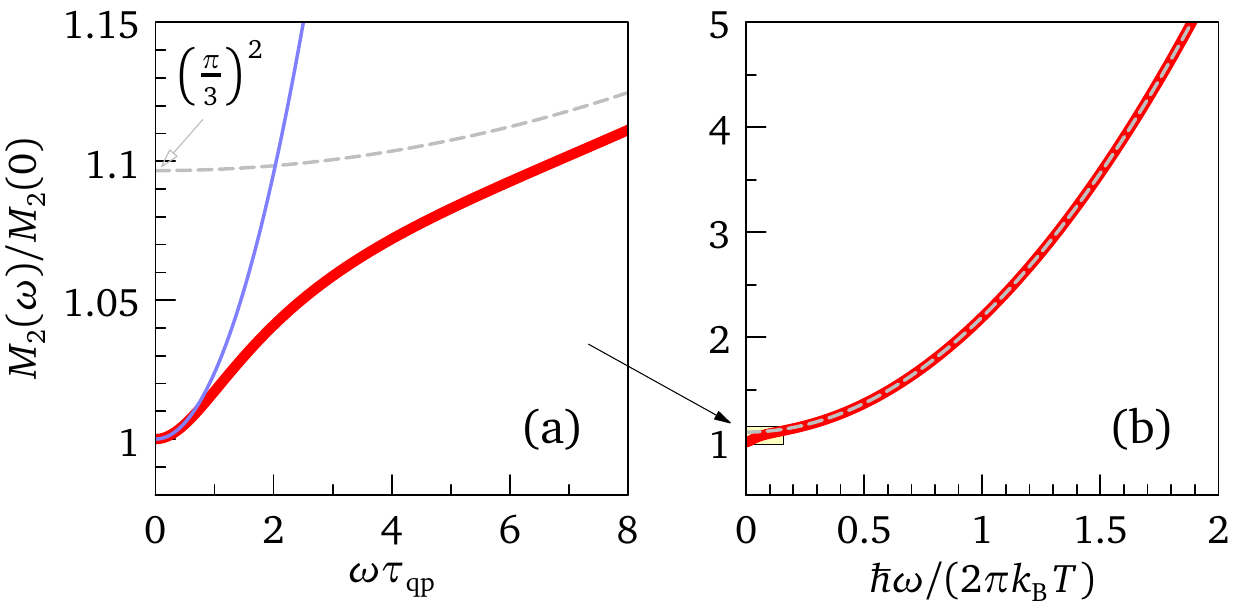}
\caption{\label{fig:memory}
Imaginary part of the memory function in a Fermi liquid at $T=0.02\T0$ (thick
red lines). (a) At frequencies $\omega<\tauqp^{-1}$, $M_2(\omega)$ increases as
$\omega^2$ (thin blue line), with a temperature-dependent curvature given by
Eq.~(\ref{eq:memory_Drude}). (b) In the thermal regime, $M_2(\omega)/M_2(0)$
scales as $(\pi/3)^2(1+x^2)$ with $x=\hbar\omega/(2\pi\kT)$ (dashed line).
}
\end{figure}

Finally, we plot in Fig.~\ref{fig:memory} the imaginary part of the memory
function, defined according to Eq.~(\ref{eq:conductivity_memory}) with the
conductivity given by the Fermi-liquid expressions (\ref{eq:Fermi_liquid}).
Figure~\ref{fig:memory}a displays the crossover from the Drude to the thermal
regime. $M_2(\omega)$ increases quadratically for $\omega\to0$, with a
temperature-dependent curvature given in Appendix~\ref{app:memory_function}. In
the thermal regime, the expansion (\ref{eq:Sxyinf}) leads to the following form:
	\begin{multline}\label{eq:memory_thermal}
		M(\omega)\approx\left(\frac{1}{\tilde{Z}}-1\right)\omega\\
		+i\frac{1}{\tilde{Z}}\frac{2}{3\pi\hbar\kTo}\left[
		(\hbar\omega)^2+(2\pi\kT)^2\right],
	\end{multline}
where $\tilde{Z}=Z\Phi(0)/(\epsilon_0\omega_p^2)$. The imaginary part is shown
in Fig.~\ref{fig:memory}(b). At $\omega=0$, Eq.~(\ref{eq:memory_thermal})
extrapolates to a value larger than the exact value $iM_2(0)$ by a factor
$(\pi/3)^2$ (see Appendix~\ref{app:memory_function}). Similarly, the
approximations (\ref{eq:sigma-tauopt}) and (\ref{eq:tauopt}) for the
conductivity deviate slightly from the exact conductivity in the limit
$\omega\to0$. Setting $\omega=0$ in these equations yields
$\sigma(0)=(3/\pi)^2\sdc$. All these observations are due to the fact that the
expressions (\ref{eq:tauopt}) and (\ref{eq:memory_thermal}) hold in the thermal
regime, but are not accurate at very low frequencies.

\subsection{Coherent and ``incoherent'' regimes}

\begin{figure}[tb]
\includegraphics[width=0.85\columnwidth]{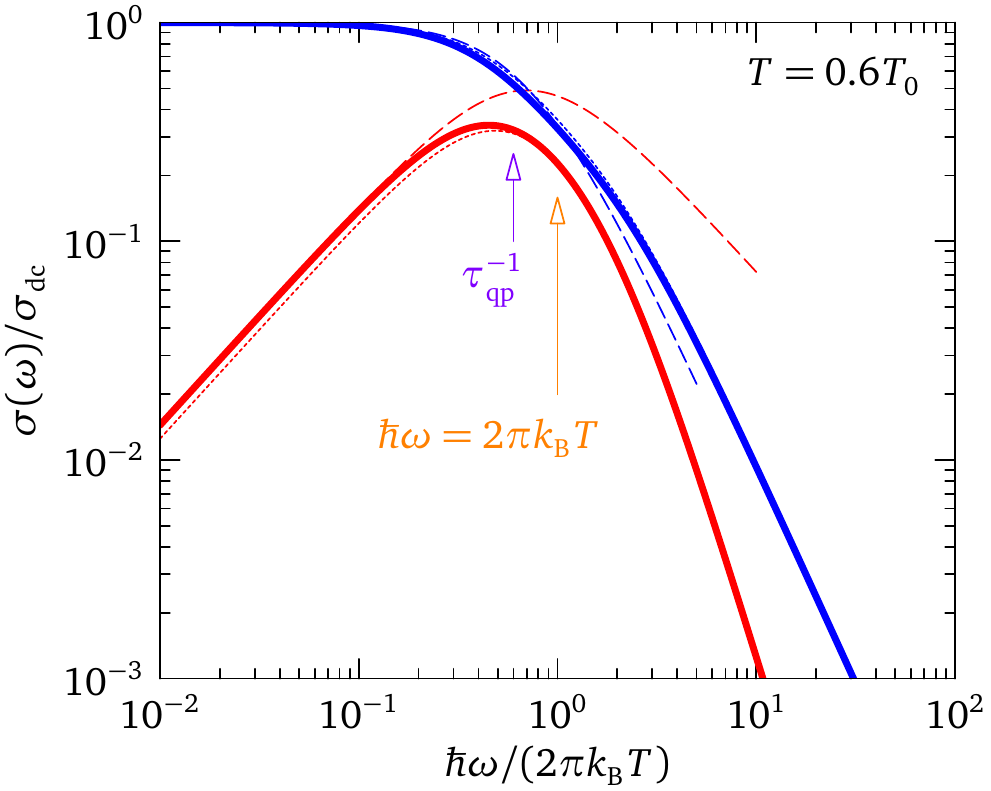}
\caption{\label{fig:log-log-highT}
Fermi-liquid conductivity Eqs.~(\ref{eq:Fermi_liquid}) at high temperature
(solid lines), above the temperature $T_1$ defined in Fig.~\ref{fig:regimes}.
The blue (red) lines show the real (imaginary) part of the conductivity. The
dashed lines show the low-frequency Drude-like behavior given by
Eq.~(\ref{eq:Drude_regime}). The dotted lines show Eq.~(\ref{eq:sigma-tauopt}).
}
\end{figure}

It is seen from Fig.~\ref{fig:regimes} that the three frequency scales
$\omega_{\text{L,\,H}}$ and $2\pi\kT/\hbar$ defined above merge when the
temperature is raised above $T_1=3\T0/8$. Above $T_1$, an ``incoherent'' regime
is found in which the thermal scale $2\pi\kT/\hbar$ becomes the only
characteristic frequency scale and $\sigma_2$ becomes smaller than $\sigma_1$ at
all frequencies, because $\tauopt^{-1}>\omega$ (see
Fig.~\ref{fig:log-log-highT}). The term incoherent is put in quotes, because at
such a high temperature $T \sim\T0$ the scaling form of the self-energy may be
no longer valid. The actual incoherent regime occurs due to the breakdown of the
Fermi liquid form (\ref{eq:self-energy}) altogether. The question whether
the scaling form applies is then irrelevant.

The reciprocal argument, on the other hand, is valid. Observing two frequencies
at which $\sigma_1=\sigma_2$ indicates that the Fermi liquid range has been
reached. On a log-log plot of $\omega\sigma_1(\omega)$ versus $\omega$, these
two crossing points coincide with two symmetric maxima located at
$\omega_{\text{L,\,H}}$, separated by a minimum at $\omega=2\pi\kT/\hbar$.

\section{Optical conductivity of a hole-doped Mott insulator within dynamical
mean-field theory}
\label{sec:OpticsDMFT}

\subsection{Model and methods} 

In this section, we present calculations of the optical conductivity for a
specific microscopic model, the single-band Hubbard model of a hole-doped Mott
insulator. The calculations are performed within single-site dynamical
mean-field theory (DMFT) \cite{georges_rmp_1996}. As we shall see, the universal
scaling form derived above allows one to identify specific signatures of
Fermi-liquid behavior in the DMFT optical conductivity, which have not been
emphasized previously. Conversely, the model calculation allows for a test of
the scaling theory, and especially of its range of validity as a function of
frequency and temperature. The optical conductivity has been calculated with
DMFT by several authors in various contexts (see, e.g.,
Refs.~\onlinecite{Jarrell-1995, Merino-2008}, and for a review and more
references, Ref.~\onlinecite{Basov-2011}). However, an explicit analysis in
connection with FL scaling laws has not been made, and such an analysis requires
calculations with very high accuracy solvers at low energy, which became
available only recently.

A semicircular density of states with half-bandwidth $D$ has been used, and the
model is considered in its paramagnetic phase. The DMFT equations were solved
using highly accurate continuous-time quantum Monte Carlo (CT-QMC)
\cite{gull_rmp_2011, TRIQS} and numerical renormalization group
\cite{bulla_rmp_2008, NRGLjubljana} (NRG) solvers. The imaginary-frequency data
from Monte Carlo were analytically continued using Pad{\'e} approximants.
Excellent quantitative agreement between the data obtained using the two
techniques was achieved in a broad temperature and frequency range. At the
lowest temperatures (below $\kT/D=0.01$) and the lowest frequencies, however,
artifacts associated with the discretization of the energy mesh and spectral
broadening become visible in the NRG data, and the CT-QMC+Pad{\'e} method
becomes preferable. Conversely, the CT-QMC+Pad{\'e} data becomes less accurate
at higher temperatures $\kT/D>0.05$, especially for larger frequencies
$\hbar\omega/D>1$. The data used here are obtained by taking the low-frequency
part ($\hbar\omega/D<0.15$) from the calculation based on the CT-QMC+Pad{\'e}
self-energies, and the high-frequency part ($\hbar\omega/D>0.15$) from the NRG
self-energies, with perfect matching in the intermediate region.

\subsection{Self-energy, local Fermi-liquid behavior, and key temperature scales}

The analysis presented here is based on the DMFT dataset used earlier in
Ref.~\onlinecite{Deng-2012}. There, transport and thermodynamic properties were
discussed in detail, but only some aspects of the optical conductivity were
addressed. For convenience, we restate here the key temperature scales
identified in this previous work, as well as their evolution with doping. The
actual data considered in detail later in the present paper are for a doping
level $\delta=20\%$ and a coupling $U/D=4$ (at which the undoped system is a
Mott insulator), but we mention the doping evolution of the key quantities.

The quasiparticle weight $Z$ was found to be approximately equal to the doping
level $Z\approx \delta$ (more precisely, $Z=0.22$ was found for $\delta=0.2$). A
temperature scale of key importance is the Fermi-liquid temperature scale
$\TFL$. This was defined as the temperature below which the $T^2$ behavior of
the resistivity and the $\omega/T$ scaling of the self-energy apply. From the
DMFT data, this scale was identified as $k_{\text{B}}\TFL=0.05\delta D$, which
gives $0.01D$ for $\delta=0.2$, a very low-energy scale. In the same work
\cite{Deng-2012}, the quasiparticle features in the spectra were shown to
persist to a much higher temperature, $\delta D$. Important particle-hole
asymmetry was found in many physical properties. At a scale approximately equal
to $\delta D$, these ``resilient quasiparticles'' disappear and the system
becomes an incoherent ``bad metal'' with a resistivity that becomes larger than
the Mott-Ioffe-Regel value. It should be emphasized that the Fermi-liquid scale
$\TFL$ is one order of magnitude smaller than the scale at which the crossover
into the bad-metal regime occurs. The latter corresponds quite accurately to the
Brinkman-Rice scale $\sim\delta D$, which corresponds to the renormalized
kinetic energy of the quasiparticles and is much larger than $\TFL$ (although
both scales are proportional to the doping level).
 
\begin{figure}[tb]
\includegraphics[width=\columnwidth]{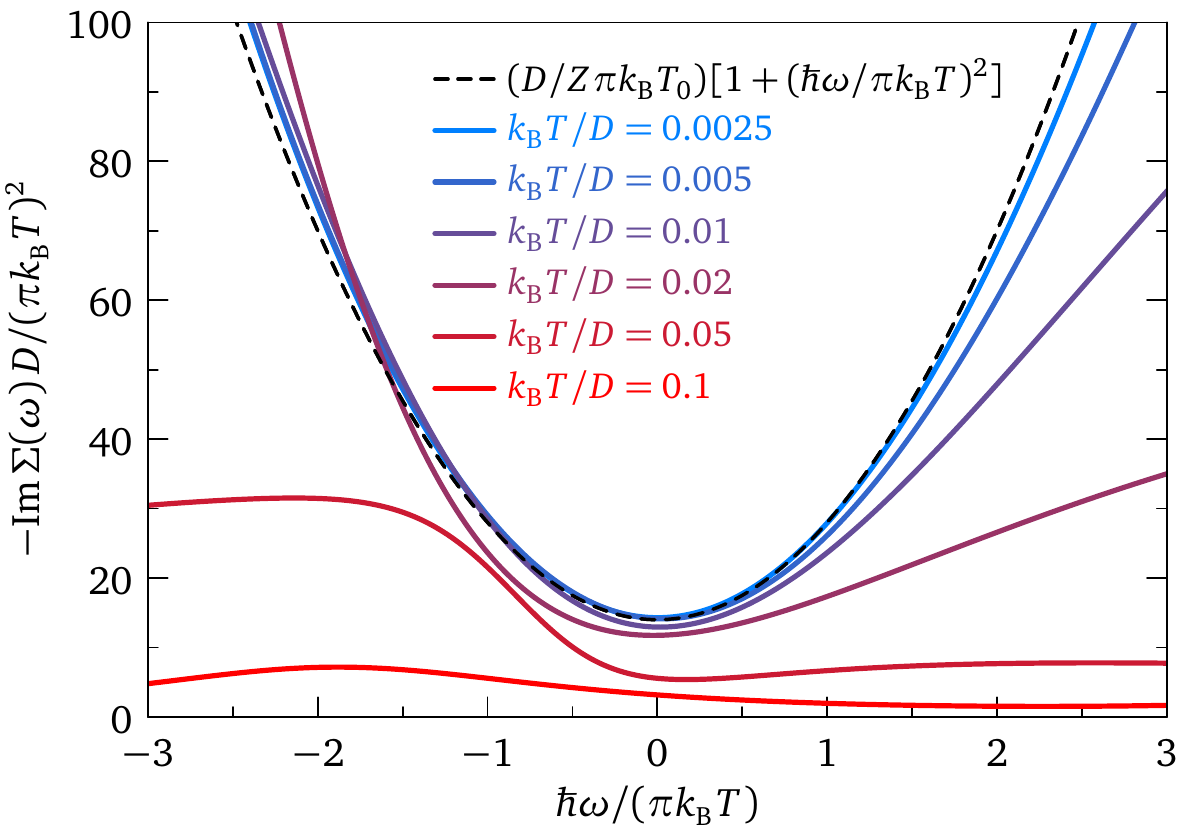}
\caption{\label{fig:self-energy-DMFT}
DMFT self-energy scaling plot. At low temperatures, the curves collapse to a
parabola. By comparing with Eq.~(\ref{eq:self-energy}) and taking into account
that one has $Z=0.22$ for a doping $\delta=0.2$, one can determine $\kTo\approx
0.1D \approx 10k_{\text{B}}\TFL$.
}
\end{figure}

In Fig.~\ref{fig:self-energy-DMFT}, we display the imaginary part of the
self-energies $\Sigma$, as a scaling plot
$-\text{Im}\,\Sigma(\omega,T)D/(\pi\kT)^2$ versus $\hbar\omega/(\pi\kT)$. The
data nicely obey the Fermi-liquid law $\propto 1+(\hbar\omega/\pi\kT)^2$ at low
temperatures. By comparing with the prefactor of this scaling behavior as
defined in Eq.~(\ref{eq:self-energy}), one determines the scale $\T0$ to be (at
$\delta=20\%$): $\kTo \approx 0.1D$, so that $\T0\approx 10\TFL$. For an
arbitrary doping level, one finds, $\kTo\approx 0.57 \delta D$. The scale $\T0$
is thus rather close in magnitude to the Brinkman-Rice scale, while $\TFL$ is an
order of magnitude smaller. Therefore, when analyzing the DMFT results in the
light of Fig.~\ref{fig:regimes} and of the scaling analysis, it should be
remembered that FL behavior actually fully applies only below $\TFL=0.1\T0$.

At higher temperatures, deviations from FL rapidly appear for electron-like
($\omega>0$) single-particle excitations (see Fig.~\ref{fig:self-energy-DMFT}).
There, the deviations from the parabolic form become substantial at a frequency
$\hbar\omegae\approx\pi\TFL$, where the real part of the self-energy (not shown,
see Ref.~\onlinecite{Deng-2012}) displays a kink. For hole like excitations
($\omega<0$), the parabolic behavior is more robust, and the kink appears only
at $\omegah\approx 0.2D$. The transport (resistivity, thermopower) probes a
frequency window of a few $\kT$, and these quantities deviate from the FL
universal behavior when $\kT>\hbar\omegae$.

\subsection{Optical conductivity at low temperature}

\begin{figure}[b]
\includegraphics[width=\columnwidth]{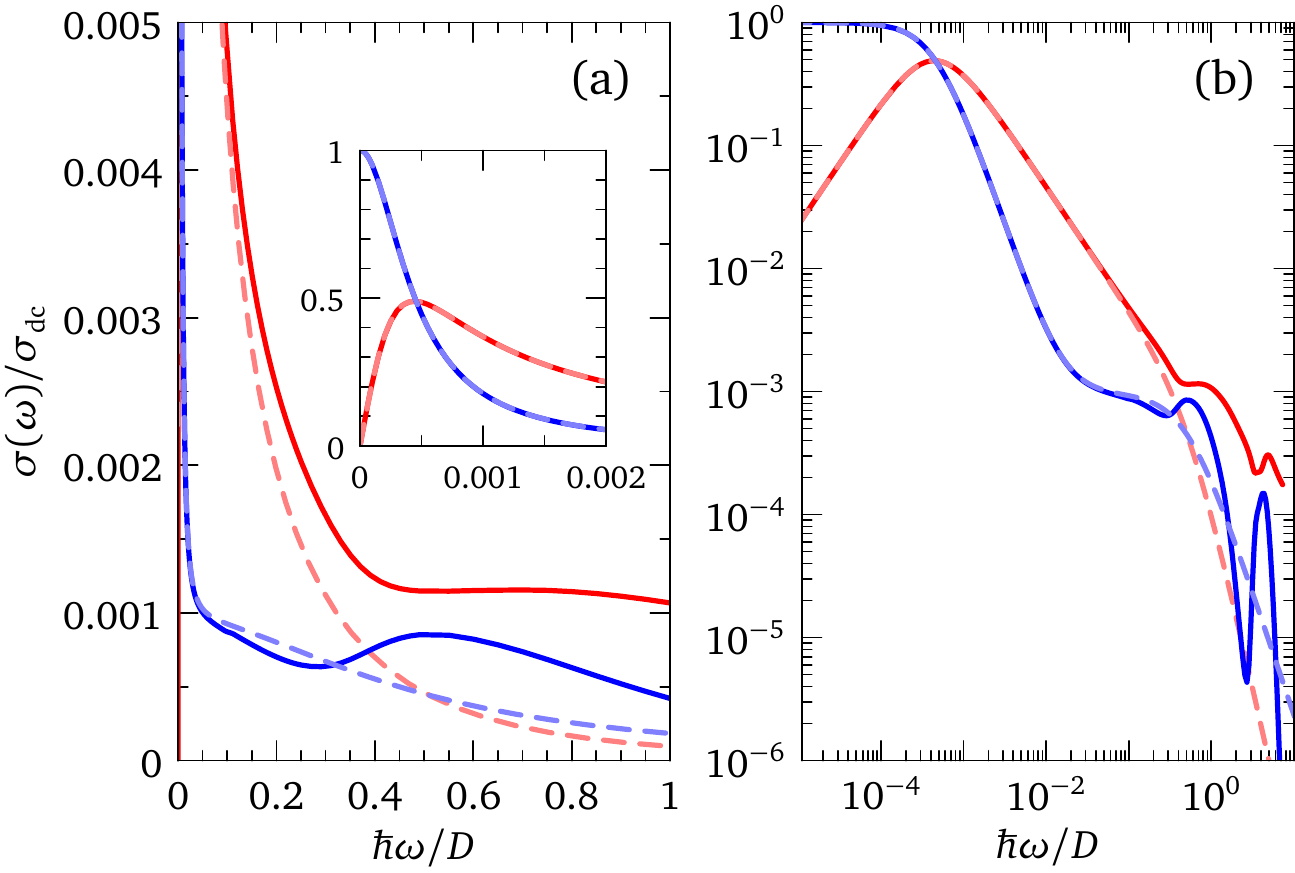}
\caption{\label{fig:conductivity1-DMFT}
Comparison of the optical conductivity at $\kT/D=0.0025D$ ($T=\TFL/4$) for the
doped Hubbard model calculated within DMFT (solid lines) to the universal FL
scaling form (dashed lines). The real (blue) and imaginary (red) parts are
plotted on a lin-lin (a) and log-log (b) scale. (Inset) Same data as in (a),
showing the Drude-like response in the low-frequency region.
}
\end{figure}

Figure~\ref{fig:conductivity1-DMFT} for $T=\TFL/4=\T0/40$ demonstrates that the
optical conductivity of the hole-doped Hubbard model obtained from DMFT is very
well described by the universal FL scaling form (\ref{eq:Fermi_liquid}) derived
in the previous section, in the low-frequency and low-temperature regimes. This
is indeed expected from the previous figure demonstrating FL scaling of the
single-particle self-energy.

When looking at $\sigma_1(\omega)$ on a lin-lin scale [see
Fig.~\ref{fig:conductivity1-DMFT}(a)], the narrow Drude peak at low-frequency is
followed at higher frequency $\hbar\omega\sim 2\pi\kT$ by a characteristic
non-Drude `foot'. On a log-log plot [see Fig.~\ref{fig:conductivity1-DMFT}(b)],
this appears as a shoulder. This does not signal non-Fermi liquid physics, as
one might naively think, but is actually a key feature of FL theory, which
signals the onset of the thermal regime at $\hbar\omega\sim 2\pi\kT$. The
non-Drude foot (or shoulder) is thus, somewhat counterintuitively, a striking
signature of Fermi-liquid behavior.

At higher frequencies, the DMFT data display two peaks. The first one, for
$\hbar\omega\approx 0.5 D$ (corresponding typically to the mid-infrared MIR
regime) is associated with the transitions between the quasiparticle band and
the lower Hubbard band. The high-frequency peak at $\hbar\omega \approx 4D\sim
U$ corresponds to the transitions to the upper Hubbard band. These peaks are,
obviously, not present in the FL expressions. Likewise, at the highest
frequencies, the bare particle dynamics (with $\sigma_2\propto 1/\omega$) is
recovered in the DMFT data, whereas extrapolating FL behavior to infinite
frequencies would lead to the incorrect behavior $\sigma_2\propto 1/\omega^3$.

\subsection{Memory function}

\begin{figure}[b]
\includegraphics[width=\columnwidth]{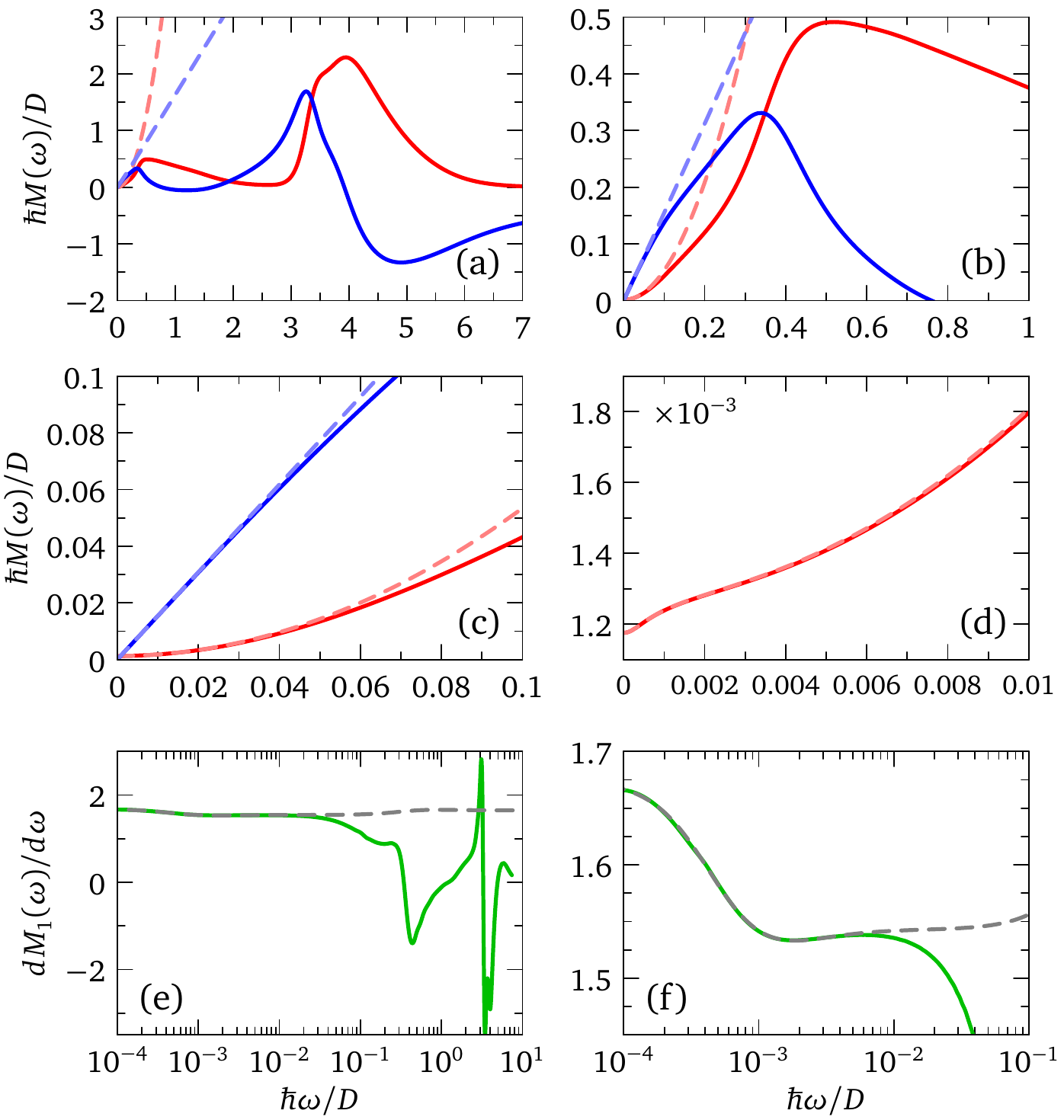}
\caption{\label{fig:memory1-DMFT}
Memory function at $\kT/D=0.0025$ ($T=\TFL/4$). (a)--(d) Real (solid blue) and
imaginary (solid red) parts of the memory function, compared with the FL scaling
forms (dashed). The data are shown for several frequency windows ranging from a
very broad one (a), to a very narrow one (d). (e) and (f\hspace{0.1em})
Frequency derivative of $M_1$ (solid lines) compared with
the FL expressions (dashed lines).
}
\end{figure}

More subtle corrections to the FL are seen in the DMFT data, also at frequencies
smaller than the MIR, in the foot-shoulder region (but above the low-frequency
kink of the self-energy). To resolve them more clearly, it is convenient to look
at the memory function. In Fig.~\ref{fig:memory1-DMFT}, we plot the memory
function at a low temperature $T=\TFL/4$.

The memory function has features on crossing the Hubbard bands [see
Fig.~\ref{fig:memory1-DMFT}(a)], which we will not discuss here. Below the MIR
scale [see Fig.~\ref{fig:memory1-DMFT}(b)], approximately linear and quadratic
behaviors are seen for the real and imaginary parts, respectively [see
Fig.~\ref{fig:memory1-DMFT}(c)]. On zooming up further [see
Fig.~\ref{fig:memory1-DMFT}(d)], one sees an excellent agreement between the
DMFT results and the FL scaling form, including the small crossover at the
lowest frequency $\omega =1/\tauqp\approx 0.001$.

Whereas agreement between the DMFT data and the FL scaling form is perfect at
the lowest frequencies, some deviations appear at a small but well-defined and
important FL frequency scale, the frequency $\omegae\approx 0.03$ associated
with the positive-energy (electron-like) ``kink''. This is seen clearly in $M_2$
or the derivative $dM_1/d\omega$ [see Fig.~\ref{fig:memory1-DMFT}(e) and
\ref{fig:memory1-DMFT}(f)]. The deviation from the FL form goes into the
direction of a smaller memory function. This is due to the fact that, in the
hole-doped Hubbard model studied here, the positive frequency ``resilient
quasiparticles'' scatter less than the parabolic behavior from Landau FL theory
would predict. We stress that, for the hole-doped Hubbard model, the
corresponding deviation cannot be observed in photoemission spectroscopy, since
they appear at positive excitation energies. Optical spectroscopy, because it
probes particle-hole excitations, is thus a powerful tool that could probe the
existence of these resilient quasiparticles. Moreover, combining photoemission
with precise optical spectroscopy could be used to reveal the strong
particle-hole asymmetry of these excitations. Signatures of the low-frequency
kink from optical spectroscopy would also be very interesting.

\subsection{Temperature dependence of the optical conductivity}

\begin{figure}[b]
\includegraphics[width=0.9\columnwidth]{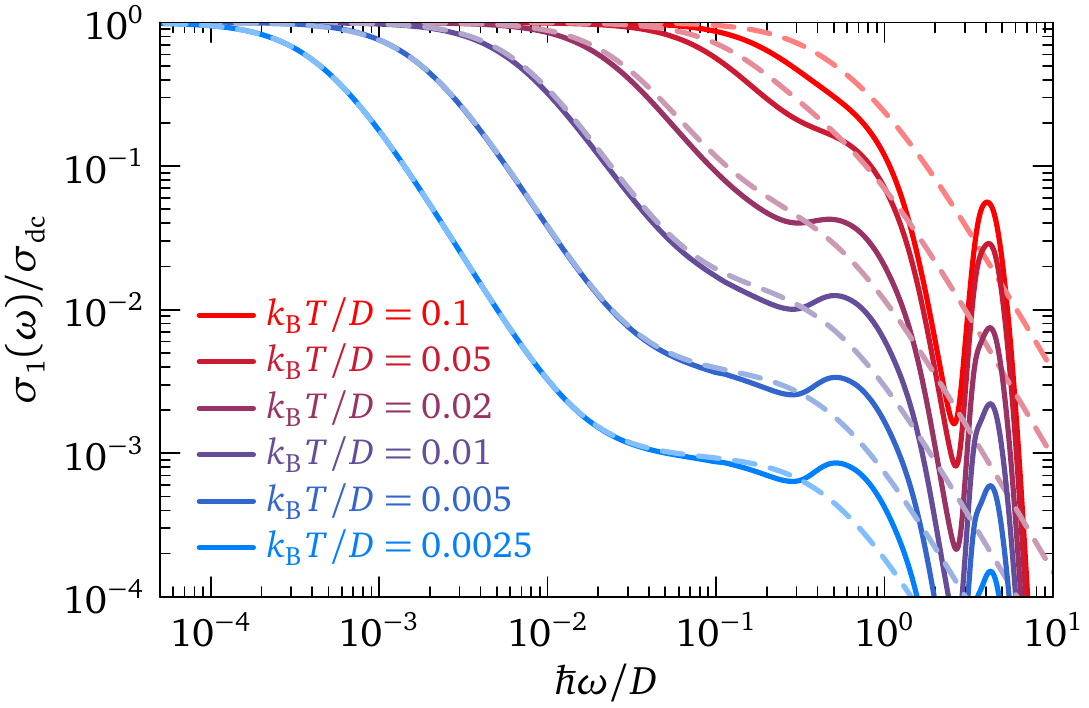}
\caption{\label{fig:conductivity2-DMFT}
Comparison of the real part of the optical conductivity in the Hubbard model
from DMFT (solid lines) with the analytical FL scaling expressions (dashed
lines). Inverse temperatures are $D/\kT=400, 200, 100, 50, 20, 10$.
}
\end{figure}

Finally, we discuss the evolution of the optical conductivity at higher
temperatures. Figure~\ref{fig:conductivity2-DMFT} displays $\sigma_1(\omega)$
for several temperatures ($D/\kT=400, 200, 100, 50, 20, 10$). The DMFT data are
compared to the analytical form dictated by the FL scaling function, in which
$\tauqp$ is determined from the FL $T^2$ dependence. The difference between the
FL and the DMFT becomes pronounced for $T\gtrsim \TFL$. Note that a milder
frequency dependence is seen above $\TFL$, both in the FL and DMFT (in the
former it occurs due to the narrowing of the thermal crossover foot-shoulder).
This warns again that interpreting apparent non-Drude power laws as a signature
of non-Fermi liquid behavior is a risky enterprise.

\begin{figure}[tb]
\includegraphics[width=\columnwidth]{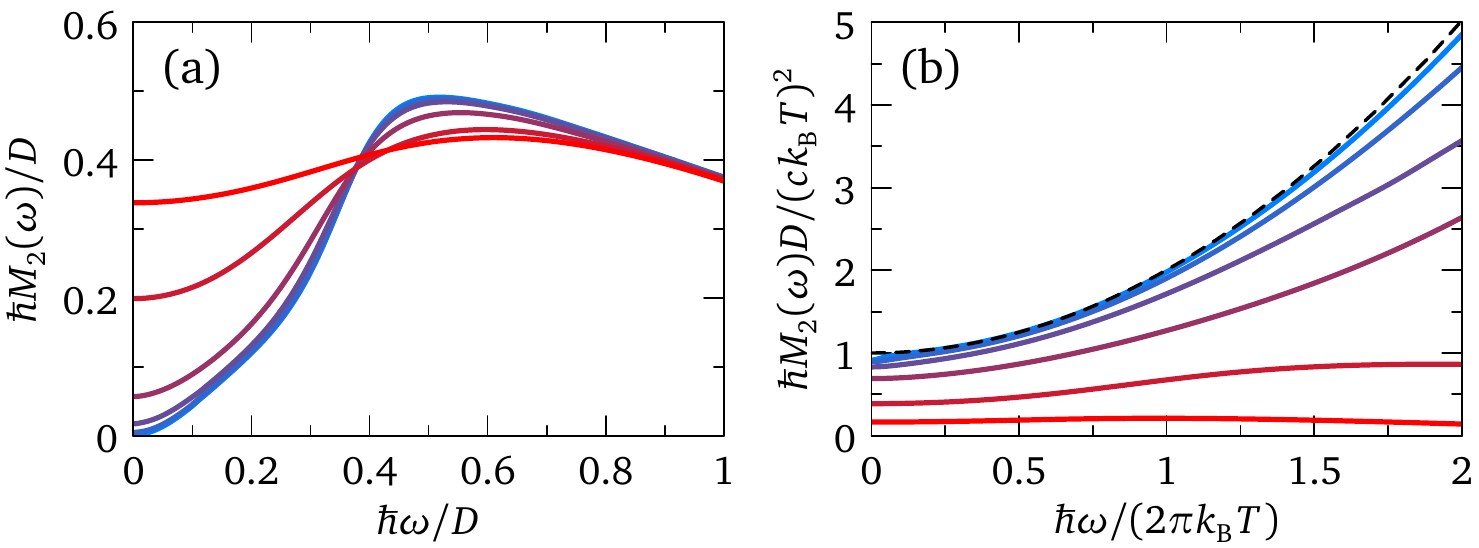}
\caption{\label{fig:memory2-DMFT}
(a) Temperature dependence of the imaginary part of the memory function in DMFT.
The temperatures and the color code are the same as in
Fig.~\ref{fig:conductivity2-DMFT}. (b) Scaling plot. The $\hbar\omega/(2\pi\kT)$
scaling applies at low temperatures and intermediate frequencies. The dashed
line shows the parabolic behavior corresponding to
Eq.~(\ref{eq:memory_thermal}).
}
\end{figure}

In Fig.~\ref{fig:memory2-DMFT}, we show the temperature evolution of the
imaginary part of the memory function. In Fig.~\ref{fig:memory2-DMFT}(b), the
data are plotted as $M_2/T^2$ versus $\omega/T$. This reveals clearly the
scaling behavior, consistent with the $(\hbar\omega)^2+(2\pi\kT)^2$ dependence
in the thermal regime of FL theory.

At low $T$, FL deviations from the quadratic dependence on frequency are seen at
the lowest frequencies, indicating the onset of the Drude regime where
$\hbar\omega\ll 2\pi\kT$. On the high-energy side, discrepancies are seen above
$\omegae$. Above $\TFL$ the data deviate from a parabola and the $\omega/T$ FL
scaling does not apply anymore.

\subsection{Relation to previous DMFT work}

The optical conductivity of doped Mott insulators  has been investigated in
several earlier single-site DMFT studies (see, e.g.,
Refs.~\onlinecite{Jarrell-1995, rozenberg_prb_1996, Merino-2000, Merino-2008,
comanac2008opticalconductivity}). The main features discussed above appearing in
the optical conductivity  (the low-frequency peak, the MIR feature associated
with transitions involving the lower Hubbard band, and the high-energy feature
associated with the upper Hubbard band) agree with these previous works. On the
other hand, the precise low-frequency dependence of the optical conductivity has
not been discussed previously. The existence of a ``thermal'' regime when
$\hbar\omega\sim \kT$, and the associated non-Drude foot in $\sigma_1(\omega)$,
which is a distinctive signature of Fermi-liquid behavior, as well as the
universal scaling form describing the low-frequency regime, were not reported in
previous literature using DMFT. The technical reason is that these can only be
revealed when using high-accuracy impurity solvers in order to produce accurate
data for the self-energy on the real-frequency axis at low temperature. Such
techniques only became available recently.

Obviously, an important open issue raised by our work is the influence of
spatial correlations beyond single-site DMFT, and their consequences for the
scaling behavior and the features pointed out here. Answering this question
requires two major steps on the methodological level. First, the momentum
dependence of the self-energy must be taken into account, for example, using
cluster extensions of DMFT. Second, one must consider the possible influence of
vertex corrections, which cannot be discarded whenever the self-energy has
momentum dependence. These are active fields of current research, which go well
beyond the simple framework and observations of the present article. Some recent
studies have pioneered the investigation of vertex corrections to the optical
conductivity, such as Refs.~\onlinecite{lin_gull_millis_2009,
lin_gull_millis_prb_2010, bergeron_tremblay_prb_2011}. To what extent the
results reported here survive in the presence of strong spatial correlations is
an interesting and challenging open problem. It will demand more work on the
methodological and technical sides to provide accurate access to the
low-temperature, low-frequency regimes of interest to our study. In this
respect, let us remark that the studies based on exact diagonalization and
related approaches do not provide access to this regime (see, e.g.,
Ref.~\onlinecite{zemljic_prelovsek_prb_2005}), since they apply to finite-size
systems and hence have a limited frequency resolution.

\section{Implications for experiments}
\label{sec:Experiment}

In this section, we address the implications for the experimental optical
signatures of Fermi-liquid behavior, in the light of the theoretical results
discussed in Sec.~\ref{sec:OpticsFL}. We will also discuss what has been
obtained until now in a number of materials, in particular, the heavy-fermion
materials CePd$_3$, UPd$_2$Al$_3$, and URu$_2$Si$_2$, heli-magnetic MnSi, the
organic conductor $\kappa$-(BEDT-TTF)$_2$Cu[N(CN)$_2$]Br$_x$Cl$_{1-x}$, the
doped semiconductor SrTiO$_3$, and the doped Mott-Hubbard insulator
HgBa$_2$CuO$_{4+\delta}$. We will see that the experimental optical data
published until now do not provide a sufficiently broad spectral range to
distinguish all of the optical features of a Fermi liquid. The first signature
is the foot (or shoulder), illustrated in Figs.~\ref{fig:log-log-lowT} and
\ref{fig:conductivity1-DMFT}, which marks the deviation from a low-frequency
Drude-like behavior. This feature occurs at the frequency $\omega \sim
2\pi\kT/\hbar$---in-between two frequencies where the imaginary and real parts
of the conductivity are equal---and disappears as the temperature is raised
above $T_1$ (see Figs.~\ref{fig:regimes} and \ref{fig:log-log-highT}). The
second signature is the characteristic frequency-temperature scaling of the
optical relaxation rate shown in Eq.~(\ref{eq:tauopt}), or the equivalent
behavior of the memory function given in Eq.~(\ref{eq:memory_thermal}). While
the first signature can, in principle, be observed in the raw data, the second
requires a determination of the spectral weight in order to invert the complex
conductivity. A third signature---present in the raw data---is the minimum in
$1/\sigma_1$, when plotted as a function of temperature at finite frequency (see
Fig.~\ref{fig:minimum}).

Experimentally, the most direct clue of Fermi-liquid physics is a $T^2$ law in
the resistivity. This is usually observed between a ground-state ordering
temperature $T_c$, and a scale $\TFL$ above which additional scattering
mechanisms contribute to the resistivity with a different temperature
dependence. For $T_c<T<\TFL$, Fermi-liquid signatures are expected in the
conductivity (provided $T<T_1=3\T0/8$) around the frequency $\nu=\kT/\hbar$,
which is $\sim1.3$~THz for $T=10$~K, corresponding to 44~cm$^{-1}$. The
signature may be masked by optical phonons---with energies typically above
5~meV, i.e., 1~THz or 40~cm$^{-1}$---and interband transitions. Another possible
limitation is impurity scattering, which reduces the value of $T_1$ as discussed
in Appendix~\ref{app:impurity_scattering}.

Fermi-liquid behavior of the dc resistivity has been reported in a variety of
materials. Using a bolometric direct absorption technique, Webb \textit{et al.}
\cite{Webb-1986} observed the narrow zero-frequency mode corresponding to the
Drude peak for CePd$_3$. The presence of the peak was also indicated by Awashti
\textit{et al.} \cite{Awasthi-1989} using resonant cavities for three discrete
frequencies (for an extensive review of optical properties of heavy-electron
compounds, including Fermi-liquid aspects, see Ref.~\onlinecite{Degiorgi-1999}).

\begin{figure}[tb]
\includegraphics[width=\columnwidth]{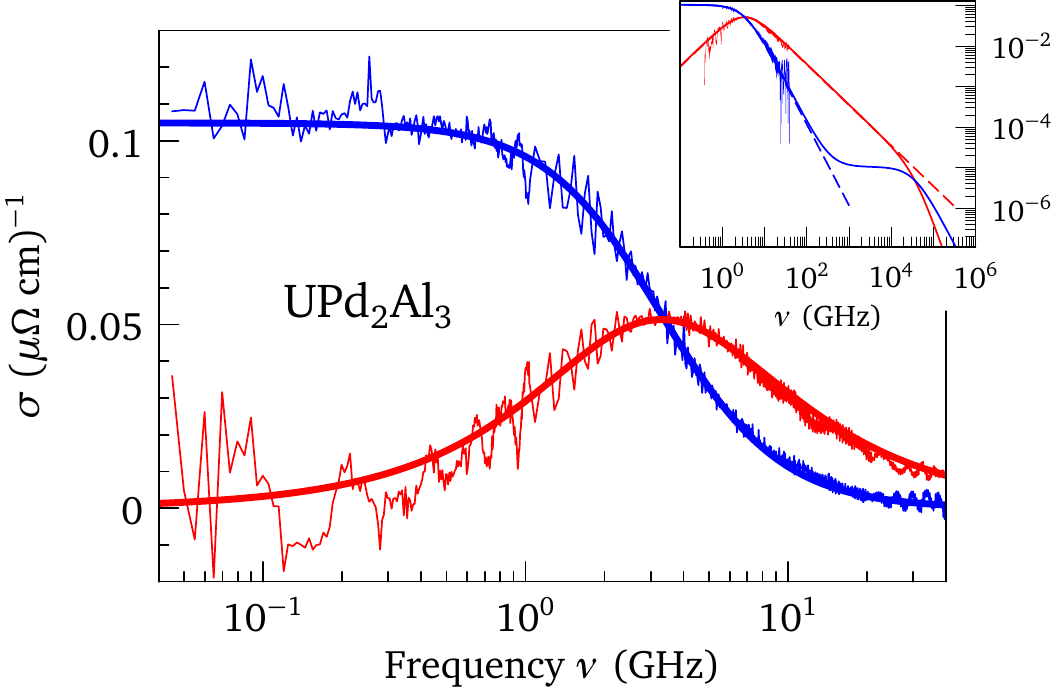}
\caption{\label{fig:UPd2Al3}
Microwave optical response of UPd$_2$Al$_3$ at $T=2.75$~K. The thin lines show
the measurements of Ref.~\onlinecite{Scheffler-2005}. The solid lines show a fit
to Eqs.~(\ref{eq:Fermi_liquid}) with the parameters $\T0=350$~K and
$Z\Phi(0)=\epsilon_0(1.6\times10^{14}\text{s}^{-1})^2$. Inset: same data on a
wider frequency range. The dashed lines show the Drude model.
}
\end{figure}

UPd$_2$Al$_3$ is a heavy-fermion material with a mass enhancement of 66
\cite{Geibel-1991}. The $5f$ moments order anti-ferromagnetically below 14.3~K
\cite{Krimmel-1992}, and a superconducting phase coexisting with the magnetic
order develops below 2~K. In the temperature range between these two
transitions, the resistivity increases as $T^2$ with a typical coefficient
$A\sim 1~\mu\Omega\text{ cm/K}^2$, suggesting Fermi-liquid behavior of the
charge carriers \cite{Ghosh-1993, Wastin-1998} with a $\TFL$ of the order of
15~K. For this material, $\sigma(\omega)$ has been measured in the range from
0.002 to 1.3 cm$^{-1}$ using a coaxial technique in Corbino geometry
\cite{Scheffler-2005}. Figure~\ref{fig:UPd2Al3} shows the microwave data of
Ref.~\onlinecite{Scheffler-2005}, measured at $T=2.75$~K, well below $\TFL$. The
conductivity can be well fitted to the Drude model,
Eq.~(\ref{eq:Drude_conductivity}), with the parameters
$\sdc=0.105~(\mu\Omega\text{ cm})^{-1}$ and $\tauD=4.7\times10^{-11}$~s. This
led the authors to the conclusion that impurity scattering dominates in this
frequency range. However, the extrapolated dc resistivity of $\sim
9~\mu\Omega\text{ cm}$ is of the order $AT^2$, which may also indicate that
electron-electron scattering dominates the dc resistivity. As a matter of fact,
the data can be equally well fitted to the Fermi-liquid formula
(\ref{eq:Fermi_liquid}), as show in Fig.~\ref{fig:UPd2Al3}. This provides an
alternative interpretation for the success of the Drude model: the scattering
rate is due to electron-electron interactions, but it is dominated by the
temperature in this low-frequency range. At the experimental temperature
$T=2.75$~K, the thermal regime where deviations from the Drude model due to the
frequency dependence of the scattering rate are expected, is around
$\kT/\hbar=360$~GHz (inset of Fig.~\ref{fig:UPd2Al3}). Note that again $\TFL$ is
an order of magnitude smaller than the value $\T0\sim 350$~K resulting from the
fit in Fig.~\ref{fig:UPd2Al3}.

Recently, Nagel \textit{et al.}\ \cite{Nagel-2012} reported optical data for
URu$_2$Si$_2$, exhibiting the $1/\tauopt(\omega,T)\propto (\hbar\omega)^2
+(p\pi\kT)^2$ dependence of the optical scattering rate with $p\approx 1$, i.e.\
well below $p=2$ expected for a Fermi liquid (see Eq.~\ref{eq:tauopt}). To our
knowledge, the two cases in which a value of $p$ closest to 2 has been observed
are the $\kappa$-(BEDT-TTF)$_2$Cu[N(CN)$_2$]Br$_x$Cl$_{1-x}$ organic compound
($p\approx2.4$) \cite{Dumm-2009} and the underdoped HgBa$_2$CuO$_{4+\delta}$
($p\approx1.5$) \cite{Mirzaei-2012}. The optical conductivity of
$\kappa$-(BEDT-TTF)$_2$Cu[N(CN)$_2$]Br$_x$Cl$_{1-x}$ does have a narrow Drude
peak followed by a foot. However, the frequency region $\omega<\omegaL$ has not
been fully explored.

For most Fourier-transform spectrometers, the lower limit of the spectrometer
range is about 20--40~cm$^{-1}$. As a result, experiments typically see only the
upper frequency part of the shape shown in Fig.~\ref{fig:log-log-lowT},
containing the characteristic frequency scales $\omegaH$, while the part of the
spectrum around $\omegaL$ is usually not reported. This upper frequency part is
itself often approximated by a Drude form, which, however, should not be
confused with the true low-frequency behavior of the conductivity. For a number
of materials, the experimental reports have hinted toward a crossover to a low
frequency regime that could be Fermi-liquid like. For example, in MnSi the
resistivity is found to follow $\rho=AT^2$ in the helimagnetic phase below 30~K,
while the optical conductivity is given by
$\sigma_1(\omega)\propto\omega^{-0.5}$ down to 30~cm$^{-1}$, even for data taken
at 10~K, indicating that a crossover to Fermi-liquid like behavior, if present,
would have to occur below 30~cm$^{-1}$ (see Ref.~\onlinecite{Mena-2003}).

This is remedied to a large extent by state of the art time-domain terahertz
spectrometers, which typically span a frequency range from 1 up to
100~cm$^{-1}$. The limiting factor in the latter case is provided by diffraction
in the long-wavelength limit, imposed by the finite sample size, which is
typically of the order of a few squared millimeters. In Nb-doped SrTiO$_3$,
$T^2$ resistivity in a broad temperature range is an indication for Fermi-liquid
behavior \cite{Klimin-2012, vanderMarel-2011}. Time-domain terahertz data reveal
a very narrow zero-frequency mode \cite{vanMechelen-2008}, similar to that in
Fig.~\ref{fig:log-log-lowT}. On the other hand, whether or not the foot-shoulder
at higher frequency is present in the data, has been impossible to establish
because strong optical phonons mask the electronic part of the spectrum in the
relevant frequency range.

A team involving the present authors recently reported Fermi-liquid like
features for underdoped high-$T_c$ cuprates in the pseudogap phase, in
particular, the scaling collapse of the form $1/\tauopt(\omega,T)\propto
(\hbar\omega)^2 + (p\pi\kT)^2$ for a broad range of $\omega$ and $T$
\cite{Mirzaei-2012}. Further analysis of these data in relation to the features
discussed in the previous sections is under way.

The examples discussed in this section illustrate the experimental challenges
that need to be met in order to establish in a single material the different
aspects of the Fermi-liquid optical conductivity, such as the one displayed in
Fig.~\ref{fig:log-log-lowT}. To the best of our knowledge, this is not available
in the literature. Still, with appropriate choice of materials, sample geometry,
and optical instrumentation, those challenges may ultimately be met.

\section{Conclusion}
\label{sec:Conclusion}

We have derived analytical formulas for the universal scaling laws that describe
the optical response of local Fermi liquids. These laws depend on two variables,
$\hbar\omega/(2\pi\kT)$ and $\omega\tauqp=\hbar\omega/(2\pi\kT)(\T0/T)$, where
$\tauqp$ is the temperature-dependent quasiparticle relaxation time on the Fermi
surface, and $\T0$ is a characteristic temperature, above which the scattering
rate $\hbar/\tauqp$ is larger than $2\pi\kT$. In the most interesting
temperature range, $T\ll\T0$, two regimes of frequency can be distinguished:
$\hbar\omega\ll 2\pi\kT$ and $\hbar\omega\gtrsim 2\pi\kT$. In the former
low-frequency regime, the conductivity displays a Drude-like response, with a
saturation of the real part below the frequency $\tauqp^{-1}$, and a
$1/\omega^2$ decay above $\tauqp^{-1}$. In the latter so-called ``thermal''
regime, clear signatures of FL behavior are identified, which are not contained
in the simple Drude form. In particular, the Fermi liquid behaves inductively in
the thermal regime, and the real part of the conductivity changes to a frequency
dependence much weaker than $1/\omega^2$. The feature appears as a shoulder in a
log-log plot, more as a `foot' in a linear plot of the conductivity. These
scaling laws can help distinguishing among the ubiquitous deviations from
Drude-like behavior those which prove FL behavior from those which disprove it.
We have illustrated this by comparing the scaling forms with DMFT data for the
Hubbard model at low doping. Finally, we have reviewed a number of experimental
works, and concluded that, while hints of FL behavior have been reported,
several of the characteristic features associated with the thermal regime remain
to be observed.

\acknowledgments

We acknowledge useful discussions with A.~Chubukov, M.~Dressel, A. J. Leggett,
and T.~Timusk, and are most grateful to M.~Ferrero for discussions and for
sharing his codes and his expertise on CT-QMC algorithms and analytical
continuation methods. This work was supported by the Swiss National Science
Foundation through Division II and MaNEP.

\appendix

\section{Kubo formula for a local Fermi liquid}
\label{app:Kubo}

In general, the long-wavelength linear conductivity tensor is related to the
current susceptibilities by the Kubo formula
	\begin{equation}\label{eq:sigma_alphabeta}
		\sigma_{\alpha\beta}(\omega)=\frac{ie^2}{\omega}
		\left[\chi_{\alpha\beta}(\omega)
		-\chi_{\alpha\beta}(0)\right].
	\end{equation}
$\chi_{\alpha\beta}(\omega)$ is the retarded macroscopic current-current
correlation function, which may be obtained from the analytic continuation of
the corresponding imaginary-time function:
	\begin{equation}
		\chi_{\alpha\beta}(i\Omega_n)=
		-\frac{1}{L^d}\int_0^{1/\kT}d\tau\,e^{i\Omega_n\tau}
		\langle j_{\alpha}(\vec{q}=0,\tau)j_{\beta}(\vec{0},0)\rangle,
	\end{equation}
where $L$ is the system size, $d$ the dimensionality, $\vec{j}$ is the
paramagnetic current operator, and $\Omega_n=2n\pi\kT$ are the bosonic Matsubara
frequencies. The diamagnetic contribution is real and diagonal, and is commonly
expressed in terms of the carrier density $n$ and the carrier mass $m$, as
$\chi_{\alpha\beta}(0)=-\delta_{\alpha\beta}(n/m)$. In interacting Fermi
systems, the current susceptibilities can be formally represented by two classes
of diagrams. The first class describes the propagation of uncorrelated
particle-hole pairs, and can be summed to give the so-called ``particle-hole
bubble'', formulated in terms of the single-particle spectral function
$A(\vec{k},\varepsilon)$. The expression of the bubble is
	\begin{multline}\label{eq:chialphabeta}
		\chi_{\alpha\beta}(\omega)=\frac{1}{L^d}
		\sum_{\vec{k}\sigma}v^{\alpha}_{\vec{k}}v^{\beta}_{\vec{k}}
		\int_{-\infty}^{\infty} d\varepsilon_1d\varepsilon_2\,
		A(\vec{k},\varepsilon_1)A(\vec{k},\varepsilon_2)\\
		\times\frac{f(\varepsilon_1)-f(\varepsilon_2)}
		{\hbar\omega+i0^++\varepsilon_1-\varepsilon_2},
	\end{multline}
where $f(\varepsilon)$ is the Fermi function, and
$\vec{v}_{\vec{k}}=(1/\hbar)\vec{\nabla}\Ek$ is the group velocity with $\Ek$
the noninteracting electron dispersion. The second class of diagrams contains
all processes involving interactions between the particle and the hole, the
so-called ``vertex corrections'', which have been shown to vanish by symmetry in
local Fermi liquids characterized by a momentum-independent self-energy
\cite{Khurana-1990}. This represents a considerable simplification, since the
exact current-current correlation function reduces to
Eq.~(\ref{eq:chialphabeta}). For a local self-energy
$\Sigma(\varepsilon)=\Sigma_1(\varepsilon)+i\Sigma_2(\varepsilon)$, the spectral
function is
	\begin{equation}\label{eq:A_k_e}
		A(\vec{k},\varepsilon)=\frac{-\Sigma_2(\varepsilon)/\pi}
		{[\varepsilon-\xi_{\vec{k}}-
		\Sigma_1(\varepsilon)]^2+[\Sigma_2(\varepsilon)]^2},
	\end{equation}
with $\xi_{\vec{k}}=\Ek-\mu$ and $\mu$ is the chemical potential. Inserting
Eq.~(\ref{eq:A_k_e}) into Eq.~(\ref{eq:chialphabeta}), taking the real part in
Eq.~(\ref{eq:sigma_alphabeta}), and defining the isotropic conductivity as an
average over the spatial coordinates,
$\sigma(\omega)\equiv(1/d)\sum_{\alpha=1}^d\sigma_{\alpha\alpha}(\omega)$,
directly leads to Eqs.~(\ref{eq:sigma_Phi}).

We note in passing that Eqs.~(\ref{eq:sigma_Phi}) provides an interesting
generalization of the Drude formula for the dc conductivity: $\sigma_1(0)$ can
be transformed into
	\begin{multline}
		\sigma_1(0)=\int_{-\infty}^{\infty} d\varepsilon\,[-f'(\varepsilon)]
		\frac{\hbar}{2|\Sigma_2(\varepsilon)|}\\ \times
		\int_{-\infty}^{\infty} du\,\Phi\Big(\varepsilon-\Sigma_1(\varepsilon)-
		u|\Sigma_2(\varepsilon)|\Big)\frac{2/\pi}{(u^2+1)^2},
	\end{multline}
with $f'$ the derivative of the Fermi function. Since $\Phi$ generally is a slow
function of its argument, we may expand about $u=0$, and get at lowest order,
	\begin{equation}
		\sigma_1(0)\approx\int_{-\infty}^{\infty} d\varepsilon\,
		[-f'(\varepsilon)]\frac{\hbar}{2|\Sigma_2(\varepsilon)|}
		\Phi\Big(\varepsilon-\Sigma_1(\varepsilon)\Big).
	\end{equation}
If $\Phi$ is taken constant, we recover the well-known fact that the dc
conductivity does not depend on the real part of the self-energy, explaining the
absence of dynamical effective mass renormalization in the Drude formula. A weak
dependence on the dynamical effective mass only enters through the energy
dependence of the function $\Phi(\xi)$. It can also be seen that, if one expands
$\Phi(\xi)$ in Eq.~(\ref{eq:sigma_1}), the first-order contribution to
$\sigma_1(\omega)$ vanishes identically if the self-energy has the symmetry
property $\Sigma(-\varepsilon)=-\Sigma^*(\varepsilon)$, as it is the case for
the model Eq.~(\ref{eq:self-energy}).

The function $\Phi(\xi)$ in Eq.~(\ref{eq:Phi_xi}) is generally a slowly-varying
function of its argument. After replacing $\Phi(\xi)$ by $\Phi(0)$ in
Eq.~(\ref{eq:sigma_1}), the $\xi$ integration can be performed by means of the
identity
	\begin{multline}
		\frac{1}{\pi}\int_{-\infty}^{\infty} d\xi\frac{\Gamma_1\Gamma_2}
		{[(\varepsilon_1-\xi)^2+\Gamma_1^2][(\varepsilon_2-\xi)^2+\Gamma_2^2]}\\
		=\frac{\Gamma_1+\Gamma_2}
		{(\varepsilon_1-\varepsilon_2)^2+(\Gamma_1+\Gamma_2)^2},
	\end{multline}
leading to
	\begin{multline}
		\sigma_1(\omega)=\frac{\Phi(0)}{\omega}\int_{-\infty}^{\infty}
		d\varepsilon\,[f(\varepsilon)-
		f(\varepsilon+\hbar\omega)]\\\times\frac{-\Sigma_2(\varepsilon)-
		\Sigma_2(\varepsilon+\hbar\omega)}{[\hbar\omega+\Sigma_1(\varepsilon)-
		\Sigma_1(\varepsilon+\hbar\omega)]^2+[\Sigma_2(\varepsilon)+
		\Sigma_2(\varepsilon+\hbar\omega)]^2}.
	\end{multline}
Equivalently, this can be rewritten as:
	\begin{multline}\label{eq:sigma1}
		\sigma_1(\omega)=\frac{\Phi(0)}{\omega}\int_{-\infty}^{\infty}
		d\varepsilon\,[f(\varepsilon)-
		f(\varepsilon+\hbar\omega)]\\\times
		\text{Re}\left[\frac{i}{\hbar\omega+\Sigma^*(\varepsilon)-
		\Sigma(\varepsilon+\hbar\omega)}\right].
	\end{multline}
Equation~(\ref{eq:sigma1}) shows that $\sigma_1(\omega)$ is given by
Eq.~(\ref{eq:sigma_Sigma}). To prove that this is also the correct expression
for $\sigma_2(\omega)$, we will show in the remainder of this section, that
$\sigma(\omega)$ in Eq.~(\ref{eq:sigma_Sigma}) is an analytical function of the
complex variable $\omega$ in the upper half of the complex plane, and vanishes
faster than $1/\omega$ for $|\omega|\to\infty$. Any function with these two
properties satisfies Eq.~(\ref{eq:KK}), implying that Eq.~(\ref{eq:sigma_Sigma})
is the one and only function consistent with both Eqs.~(\ref{eq:KK}) and
(\ref{eq:sigma1}).

We first note that the difference of Fermi functions in
Eq.~(\ref{eq:sigma_Sigma}) is proportional to $\omega$ as $\omega\to0$. Hence
$\sigma(\omega)$ is analytic at $\omega=0$. In order to get rid of the poles
generated by the second Fermi function in Eq.~(\ref{eq:sigma_Sigma}), when
continuing $\omega$ into the complex plane, we shift the integration variable in
the second term, and rewrite the conductivity as
	\begin{multline}\label{eq:sigma-split}
		\sigma(\omega)=\frac{i\Phi(0)}{\omega}\int_{-\infty}^{\infty}
		d\varepsilon\,f(\varepsilon)\left[
		\frac{1}{\hbar\omega+\Sigma^*(\varepsilon)
		-\Sigma(\varepsilon+\hbar\omega)}\right.\\
		\left.-\frac{1}{\hbar\omega+\Sigma^*(\varepsilon-\hbar\omega)
		-\Sigma(\varepsilon)}\right].
	\end{multline}
The self-energy, like the Green's function, is analytic everywhere except
possibly on the real axis, and decays at infinity provided that an asymptotic
real value, corresponding to a shift of the chemical potential, is subtracted
out. These analytic properties mean that the self-energy obeys
	\begin{align}\label{eq:Sigma_causal}
		\nonumber
		\Sigma(z)&=\int_{-\infty}^{\infty} dE\,
		\frac{-\frac{1}{\pi}\text{Im}\,\Sigma(E)}{z-E}\\
		&=\int_{-\infty}^{\infty} dE\,(z^*-E)
		\frac{-\frac{1}{\pi}\text{Im}\,\Sigma(E)}{|z-E|^2}.
	\end{align}
Here $z$ is a complex variable, and $E$ is real. As a function of $z$, the
self-energy has a branch-cut discontinuity on the real axis. The meaning of
$\Sigma(E)$ in Eq.~(\ref{eq:Sigma_causal}) must therefore be disambiguated: when
writing $\Sigma(E)$ with a real argument $E$, we actually mean $\Sigma(E+i0^+)$,
i.e., the self-energy just above the real axis, which has the property
$\text{Im}\,\Sigma(E)<0$. Since $\Sigma(z)\sim1/z$ for $|z|\to\infty$, $\omega$
dominates as $|\omega|\to\infty$ in the denominator of both terms in
Eq.~(\ref{eq:sigma-split}). This shows that $\sigma(\omega)$ decays at infinity
faster than $1/\omega$. For a vanishing self-energy, both terms have a pole at
$\omega=0$. The analytical structure of the self-energy displaces the pole in
the lower half of the complex plane. In order to see this, we can solve
iteratively the equation giving the pole, i.e.,
$\hbar\omega=\Sigma(\varepsilon)-\Sigma^*(\varepsilon-\hbar\omega)$ and a
similar expression for the first term in Eq.~(\ref{eq:sigma-split}), starting
from the value $\omega_1=0$. From Eq.~(\ref{eq:Sigma_causal}) and the recursion
relation
$\hbar\omega_{j+1}=\Sigma(\varepsilon)-\Sigma^*(\varepsilon-\hbar\omega_j)$, we
see that
	\begin{multline}
		\text{Im}\,(\hbar\omega_{j+1})=\text{Im}\,\Sigma(\varepsilon)\\
		+\text{Im}\,(\hbar\omega_j)
		\int_{-\infty}^{\infty}dE\,\frac{-\frac{1}{\pi}\text{Im}\,\Sigma(E)}
		{|\varepsilon-\hbar\omega_j-E|^2}.
	\end{multline}
The right-hand side is strictly negative if $\text{Im}\,(\omega_j)\leqslant0$,
which means that the fixed point, if any, has $\text{Im}\,(\omega)<0$. Hence the
analytic properties of the self-energy imply that the function in the square
brackets in Eq.~(\ref{eq:sigma-split}) is analytic in the upper half of the
complex $\omega$ plane.

\section{Optical sum rule for a single band}
\label{app:sum-rule}

This Appendix provides a derivation of the sum rule given by
Eqs.~(\ref{eq:def_omegap}) and (\ref{eq:sumrule_1band}), valid for a single-band
local Fermi liquid whose conductivity is given by Eq.~(\ref{eq:sigma_Phi}). On
the imaginary-frequency axis, the bubble contribution (\ref{eq:chialphabeta}) to
the current-current correlation function takes a simple form in terms of the
Green's function $\mathscr{G}(\vec{k},i\omega_n)$, with $\omega_n=(2n+1)\pi\kT$
the fermionic Matsubara frequencies:
	\begin{multline}\label{eq:chialphabetaiOmega}
		\chi_{\alpha\beta}(i\Omega_n)=\frac{\kT}{L^d}
		\sum_{\vec{k}\sigma}v^{\alpha}_{\vec{k}}v^{\beta}_{\vec{k}}
		\sum_{i\omega_n}
		\mathscr{G}(\vec{k},i\omega_n)\mathscr{G}(\vec{k},i\omega_n+i\Omega_n).
	\end{multline}
A convergence factor $e^{i\omega_n0^+}$ is implied hereafter---if not explicitly
written---in the Matsubara-frequency sums. In the complex plane,
$\chi_{\alpha\beta}(z)$ has the same analytical structure as the self-energy in
Eq.~(\ref{eq:Sigma_causal}). We therefore have
	\begin{equation}
		\chi_{\alpha\beta}(i\Omega_n=0)=\frac{1}{\pi}
		\int_{-\infty}^{\infty}d\omega\,
		\frac{\text{Im}\,\chi_{\alpha\beta}(\omega+i0^+)}{\omega}.
	\end{equation}
Using this, and the fact that $\text{Im}\,\chi_{\alpha\beta}(\omega+i0^+)$ is an
odd function of $\omega$, we obtain from Eqs.~(\ref{eq:sigma_alphabeta}) and
(\ref{eq:chialphabetaiOmega}),
	\begin{align}\label{eq:sumrule-G2}
		\nonumber
		\int_0^{\infty}d\omega\,\sigma_1(\omega)&=-\frac{\pi e^2}{2d}
		\sum_{\alpha=1}^d\chi_{\alpha\alpha}(i\Omega_n=0)\\
		&=-\frac{\pi e^2\kT}{2dL^d}\sum_{\vec{k}\sigma}v_{\vec{k}}^2
		\sum_{i\omega_n}\mathscr{G}^2(\vec{k},i\omega_n).
	\end{align}
In a local Fermi liquid, the momentum dependence of the Green's function stems
from the bare dispersion,
$\mathscr{G}(\vec{k},i\omega_n)=1/[i\omega_n-\xi_{\vec{k}}-\Sigma(i\omega_n)]$,
hence we have the property
$\mathscr{G}^2(\vec{k},i\omega_n)=d\mathscr{G}(\vec{k},i\omega_n)/d\xi_{\vec{k}}
$. The momentum sum in Eq.~(\ref{eq:sumrule-G2}) can then be converted into an
energy integral involving
$\mathscr{G}(\xi,i\omega_n)=1/[i\omega_n-\xi-\Sigma(i\omega_n)]$ as well as the
transport function $\Phi(\xi)$ defined in Eq.~(\ref{eq:Phi_xi}):
	\begin{equation}
		\int_0^{\infty}d\omega\,\sigma_1(\omega)=
		-\frac{\pi}{2}\kT\sum_{i\omega_n}\int_{-\infty}^{\infty} d\xi\,
		\Phi(\xi)\frac{d\mathscr{G}(\xi,i\omega_n)}{d\xi}.
	\end{equation}
The Matsubara-frequency sum of the Green's function gives the momentum
distribution
$\kT\sum_{i\omega_n}\mathscr{G}(\xi,i\omega_n)e^{i\omega_n0^+}=\int d\varepsilon
f(\varepsilon)\,A(\xi,\varepsilon)\equiv n(\xi)$. Together with
Eq.~(\ref{eq:def_omegap}), this proves the first relation in
Eq.~(\ref{eq:sumrule_1band}). The second relation may be obtained by noting that
$v_{\vec{k}}^2\mathscr{G}^2(\vec{k},i\omega_n)=(1/\hbar^2)[\vec{\nabla}_{\vec{k}
}\mathscr{G}(\vec{k},i\omega_n)]\cdot\vec{\nabla}_{\vec{k}}\xi_{\vec{k}}$.
Substituting this in Eq.~(\ref{eq:sumrule-G2}), and performing the momentum
integration by parts, we arrive at
	\begin{equation}\label{}
		\int_0^{\infty}d\omega\,\sigma_1(\omega)=\frac{\pi e^2}{d\hbar^2}
		\frac{1}{L^d}\sum_{\vec{k}}n(\xi_{\vec{k}})\nabla^2\xi_{\vec{k}},
	\end{equation}
which leads to the second relation in Eq.~(\ref{eq:sumrule_1band}).

\section{\boldmath Calculation of the $\S$ function}
\label{app:S}

\begin{figure}[tb]
\includegraphics[width=0.8\columnwidth]{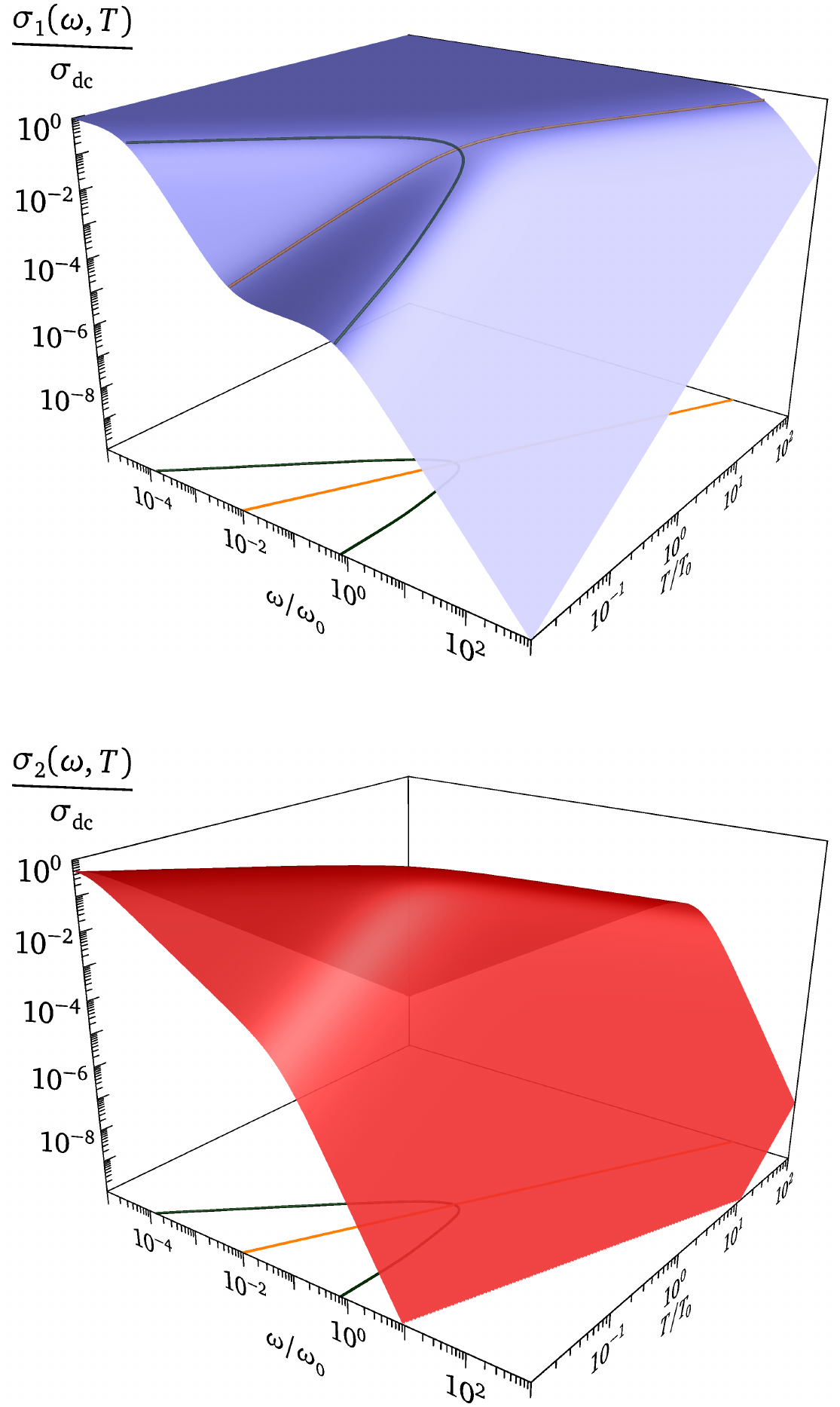}
\caption{\label{fig:S_function}
Real part (top) and imaginary part (bottom) of the optical conductivity
normalized by the dc value, as a function of frequency and temperature. $\T0$ is
defined in Eq.~(\ref{eq:self-energy}), and $\omega_0=2\pi\kTo/\hbar$. The
straight orange lines indicate $\hbar\omega=2\pi\kT$, and the green lines
corresponds to the dome in Fig.~\ref{fig:regimes}.
}
\end{figure}

In order to evaluate the $\S$ function, we use the exact representation
	\begin{multline}
		\frac{1}{e^{\pi(u-x)}+1}-\frac{1}{e^{\pi(u+x)}+1}=\\
		\frac{1}{\pi}\sum_{n=-\infty}^{+\infty}
		\left(\frac{1}{ip_n+x-u}-\frac{1}{ip_n-x-u}\right)
	\end{multline}
with $p_n=2n+1$. Changing variable in the integral in
Eq.~(\ref{eq:Fermi_liquid}), we see that each term with $p_n<0$ is equal to the
corresponding term with $p_{-n-1}=-p_n>0$, so that the sum can be rewritten as a
sum on $n\geqslant 0$. The integrations are elementary,
	\begin{equation*}
		\frac{1}{\pi}\int_{-\infty}^{\infty}du\,\left(\frac{1}{ip_n+x-u}
		-\frac{1}{ip_n-x-u}\right)\frac{1}{1+x^2-iy+u^2}\\[-0.5cm]
	\end{equation*}
	\begin{align}
		\nonumber
		&=\frac{i}{r(x,y)}\left(\frac{1}{p_n+r(x,y)+ix}-
		\frac{1}{p_n+r(x,y)-ix}\right)\\[0.2cm]
		\label{eq:definition_s}
		r(x,y)&=\sqrt{1+x^2-iy},
	\end{align}
and we may recast the sums in terms of the digamma function
$\psi(z)=\lim_{M\to\infty}\big[\ln M-\sum_{n=0}^{M}1/(n+z)\big]$, to finally
get Eq.~(\ref{eq:S_function}).

Equation~(\ref{eq:conductivity_dimensionless}) shows that the conductivity
normalized to the dc value is given by
$\S\left(\frac{\omega/\omega_0}{T/\T0},\frac{\omega/\omega_0}{(T/\T0)^2}\right)$.
This is displayed as a function of frequency and temperature in
Fig.~\ref{fig:S_function}. The function $\S$ is well approximated by the simple
form $\S(x,y)\approx1/(1+x^2-3iy/4)$ over the whole $(x,y)$ plane.

\section{Comparing Fermi-liquid and Drude formula}
\label{sec:comparison_Drude}

\begin{figure}[tb]
\includegraphics[width=\columnwidth]{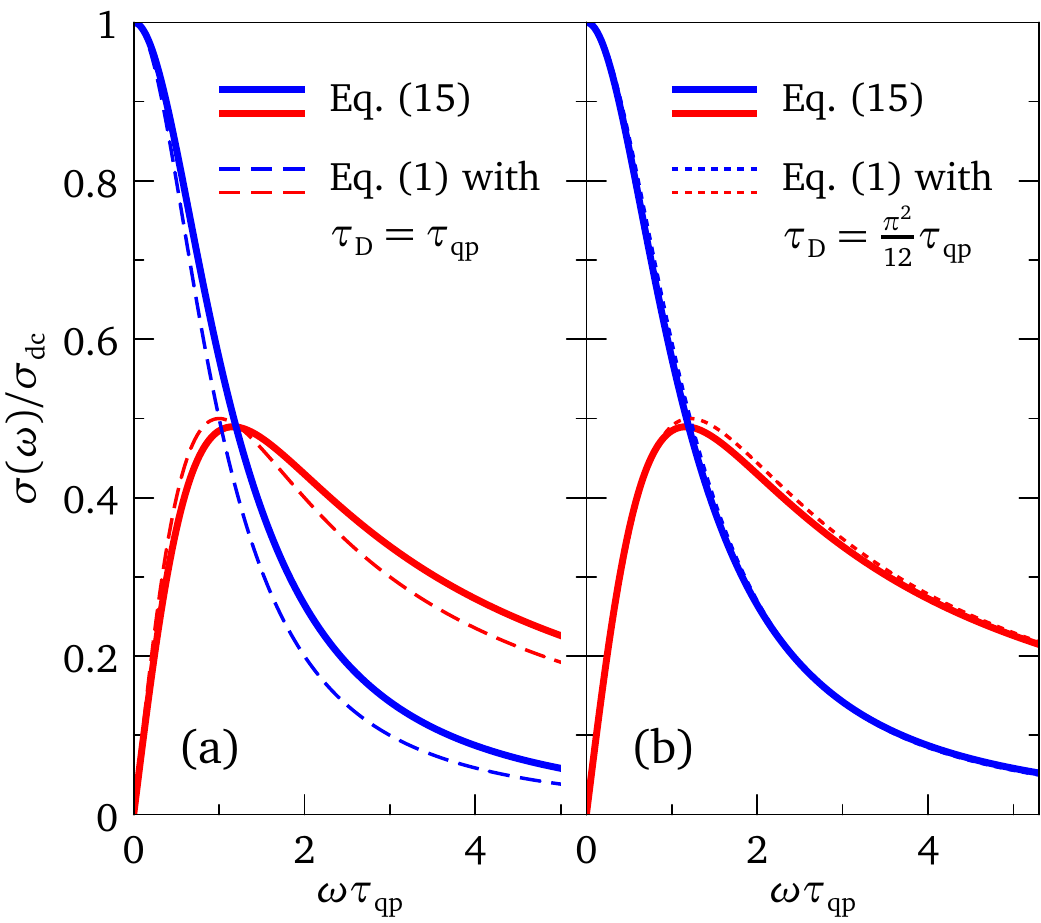}
\caption{\label{fig:comparison_Drude}
Comparison of the Fermi-liquid conductivity in the Drude regime,
Eq.~(\ref{eq:Drude_regime}), with the Drude model,
Eq.~(\ref{eq:Drude_conductivity}). The solid blue and red lines show the real
and imaginary parts of Eq.~(\ref{eq:Drude_regime}), respectively. The dashed
lines in (a) show the Drude model assuming $\tauD=\tauqp$. The dotted lines in
(b) show the Drude model assuming $\tauD=(\pi^2/12)\tauqp$ (see text).
}
\end{figure}

In the Drude regime, defined by $\hbar\omega\ll2\pi\kT$ and $\omega\tauqp\sim1$,
the conductivity of a Fermi liquid is a universal function of $\omega\tauqp$
given by Eq.~(\ref{eq:Drude_regime}). This function is displayed in
Fig.~\ref{fig:comparison_Drude}. The imaginary part $\sigma_2$ increases first
linearly with a slope $1/2+3\zeta(3)/\pi^2\approx0.865$, reaches a maximum at
$\omega\tauqp\approx 1.162$, and then decreases as
$12/\pi^2(\omega\tauqp)^{-1}$. The real part decreases as
$16/\pi^2(\omega\tauqp)^{-2}$, and crosses the imaginary part at
$\omega\tauqp\approx1.198$, slightly above the point where $\sigma_2$ has its
maximum. In the Drude model (\ref{eq:Drude_conductivity}), $\sigma_2$ increases
first like $\omega\tauD$, reaches a maximum at $\omega\tauD=1$, and then
decreases as $(\omega\tauqp)^{-1}$; $\sigma_1$ decreases as
$(\omega\tauD)^{-2}$, and crosses $\sigma_2$ at its maximum.

When approximating the low-frequency response of a Fermi liquid by the Drude
model, it is not possible to choose the time $\tauD$ such that the four
remarkable features---initial slope and maximum of $\sigma_2$, asymptotic decay
of $\sigma_1$ and $\sigma_2$---are all described exactly. If one takes
$\tauD=\tauqp$, none of these features is correctly reproduced, as illustrated
in Fig.~\ref{fig:comparison_Drude}(a). Matching the initial slope of $\sigma_2$
in both models implies $\tauD=0.865\tauqp$, matching the maximum of $\sigma_2$
means $\tauD=\tauqp/1.162$, matching the asymptotic decays of $\sigma_1$ and
$\sigma_2$ leads respectively to $\tauD=(\pi/4)\tauqp$ and
$\tauD=(\pi^2/12)\tauqp$. All four determinations of $\tauD/\tauqp$ are close to
$0.8$. The choice $\tauD=(\pi^2/12)\tauqp$, because it gives the correct
asymptotic decay of $\sigma_2$, also ensures that the spectral weight is the
same in both models by the Kramers-Kronig relations. It is furthermore the
closest to the average of the four determinations.
Figure~\ref{fig:comparison_Drude}b shows a comparison of the Fermi-liquid model
with the Drude model, assuming $\tauD=(\pi^2/12)\tauqp$.

\section{Memory function in the thermal and Drude regimes}
\label{app:memory_function}

By solving Eq.~(\ref{eq:conductivity_memory}) for $M(\omega)$, identifying
$\sigma(\omega)$ with the Fermi-liquid result (\ref{eq:Fermi_liquid}), and
expanding for large $y=\omega\tauqp$ with the help of Eq.~(\ref{eq:Sxyinf}), one
arrives at the expression (\ref{eq:memory_thermal}) for the memory function in
the thermal regime. Inserting Eq.~(\ref{eq:memory_thermal}) back into
Eq.~(\ref{eq:conductivity_memory}), one recovers the generalized Drude form of
Eqs.~(\ref{eq:sigma-tauopt}) and (\ref{eq:tauopt}). From an experimental
perspective, one sees in Eq.~(\ref{eq:memory_thermal}) that a plot of $\omega
M_2(\omega)/[\omega+M_1(\omega)]$ in the region $\hbar\omega\sim2\pi\kT$ should
allow to determine the parameter $\T0$.

We note that, while the real part of $M$ depends on the total spectral weight
$\epsilon_0\omega_p^2$, the normalized imaginary part $M_2(\omega)/M_2(0)$ does
not, since $M_2(\omega)/M_2(0)=\text{Re}\,\sdc/\sigma(\omega)$ as shown, e.g., in
Eq.~(\ref{eq:M2}). Unlike $M_1(\omega)$, $M_2(\omega)/M_2(0)$ can therefore be
determined accurately using a low-energy model. With the Fermi-liquid model
(\ref{eq:Fermi_liquid}), we have
	\begin{equation}
		\frac{M_2(\omega)}{M_2(0)}=\text{Re}\,\frac{1}{\S(x,y)}.
	\end{equation}
As $\lim_{y\to\infty}\text{Re}[1/\S(x,y)]=(\pi/3)^2(1+x^2)$, one sees that the
scaling function for $M_2(\omega)$ in the thermal regime extrapolates to the
value $(\pi/3)^2M_2(0)$ at $\omega=0$, as can be seen in Fig.~\ref{fig:memory}.

The behavior of $M_2(\omega)/M_2(0)$ in the Drude regime $\omega\tauqp<1$ is
quadratic and given by
	\begin{equation}\label{eq:memory_Drude}
		\frac{M_2(\omega)}{M_2(0)}=1+\left(\omega\tauqp\right)^2
		\left[a+b\left(\frac{T}{\T0}\right)^2\right],
	\end{equation}
where $a$ and $b$ are the numerical constants
$a=1/8+\pi^2/80-3\zeta(3)/(4\pi^2)-(3\zeta(3)/\pi^2)^2\approx0.0235$ and
$b=1/2+\pi^2/60+3\zeta(3)/\pi^2\approx1.03$. This is plotted in
Fig.~\ref{fig:memory}(a) as the thin line.

\section{\boldmath Impurity scattering, robustness of the $2\pi$ factor}
\label{app:impurity_scattering}

In this Appendix, we show that the inclusion of a constant scattering rate
$-i\Gamma$ in the Fermi-liquid self-energy (\ref{eq:self-energy}) does not
change the scaling of the optical conductivity in the thermal regime. A
frequency-independent scattering rate can crudely describe the effect of
impurity scattering. On adding $-i\Gamma$ to the self-energy, the only change
induced in Eqs.~(\ref{eq:Fermi_liquid}) occurs in the expression of the function
$\S$: in the denominator of the integrand, $1$ is replaced by
$1+2Z\tauqp\Gamma/\hbar$. The correction term $\sim\Gamma/T^2$ is the ratio
between the impurity and the electron-electron scattering rates. This change can
be absorbed into a redefinition of the variable $y$:
	\begin{equation}
		y=\frac{\omegabar}{\Tbar^2}\to y_1+iy_2
		=\frac{\omegabar+iZ\Gamma/(\pi\kTo)}{\Tbar^2}.
	\end{equation}
The same modification made in the analytical form (\ref{eq:S_function})
yields the scaling function in the presence of impurity scattering. It should be
noted that, with this modification, the function $\S$ is not unity in the dc
limit, and therefore the dc conductivity is different from $\sdc$ as defined in
Eq.~(\ref{eq:Fermi_liquid}). In particular, the residual dc resistivity can be
evaluated as $\rho_0=2\Gamma/[\hbar\Phi(0)]$.

If $\omega\gg2Z\Gamma/\hbar$, we can expand in the thermal regime like in
Eq.~(\ref{eq:Sxyinf}), which must then be replaced by
	\begin{equation}
		\S(x,y_1\to\infty,y_2)=\frac{12i}{\pi^2y_1}
		+\frac{16}{\pi^2y_1^2}\left(1+x^2+\frac{3}{4}y_2\right).
	\end{equation}
Consequently, we find that the optical scattering rate reflects the increased
quasiparticle scattering rate (with an extra factor of two), but that the
frequency-temperature scaling is unchanged:
	\begin{equation}
		\frac{\hbar}{\tauopt(\omega)}=
		\frac{2}{3\pi\kTo}\left[(\hbar\omega)^2
		+\left(2\pi\kT\right)^2\right]+2Z\Gamma.
	\end{equation}
The impurity scattering, however, reduces the frequency-temperature domain below
the dome in Fig.~\ref{fig:regimes}, where $\tauopt(\omega)>1/\omega$. The
equation of the dome in the presence of impurity scattering is
	\begin{equation}
		T_1(\omega)=\T0\sqrt{\frac{3}{4}\left(\omegabar
		-\frac{Z\Gamma}{\pi\kTo}\right)-\omegabar^2}.
	\end{equation}
Hence the dome disappears---and consequently the shoulder in the log-log plot of
$\sigma_1(\omega)$---if $Z\Gamma>3\pi\kTo/16$.

\bibliography{bib}

\end{document}